\documentclass[aps,pre,twocolumn,superscriptaddress,longbibliography]{revtex4-2}%
\usepackage{amsfonts}
\usepackage{amsmath}
\usepackage{amssymb}
\usepackage{graphicx}
\usepackage{color}
\usepackage{amsmath}
\usepackage{float}
\usepackage{mathtools}
\usepackage{algorithm}
\usepackage{algpseudocode}
\usepackage{svg}

\usepackage[colorinlistoftodos]{todonotes}
\usepackage[colorlinks=true, allcolors=blue]{hyperref}
\usepackage{mwe}
\usepackage{epstopdf}

\newcommand \be {\begin{eqnarray}}
\newcommand \ee {\end{eqnarray}}
\newcommand \ben {\begin{eqnarray}}
\newcommand \een {\end{eqnarray}}

\newcommand \rset{\{\mathbf{r}_j\}}

\newcommand \Sk{S(|\mathbf{k}|)}

\newif\iflong
\longtrue 
\begin{document}

\title{Persistent homology and topological statistics of hyperuniform point clouds}
\date{\today}

\author{Marco Salvalaglio}
\email{marco.salvalaglio@tu-dresden.de} 
\affiliation{Institute  of Scientific Computing,  Technische  Universit\"at  Dresden,  01062  Dresden,  Germany}
\affiliation{Dresden Center for Computational Materials Science (DCMS), TU Dresden, 01062 Dresden, Germany}
\author{Dominic J. Skinner}
\affiliation{NSF-Simons Center for Quantitative Biology, Northwestern University,
2205 Tech Drive, Evanston, IL 60208, USA}
\author{J\"orn Dunkel}
\affiliation{Department of Mathematics, Massachusetts Institute of Technology,
77 Massachusetts Avenue, Cambridge, MA 01239, USA}
\author{Axel Voigt}
\affiliation{Institute  of Scientific Computing,  Technische  Universit\"at  Dresden,  01062  Dresden,  Germany}
\affiliation{Dresden Center for Computational Materials Science (DCMS), TU Dresden, 01062 Dresden, Germany}
\affiliation{Cluster of Excellence Physics of Life, Technische Universit\"at Dresden, 01062 Dresden, Germany}

\begin{abstract}
Hyperuniformity, the suppression of density fluctuations at large length scales, is observed across a wide variety of domains, from cosmology to condensed matter and biological systems. 
Although the standard definition of hyperuniformity only utilizes information at the largest scales, hyperuniform configurations have distinctive local characteristics. However, the influence of global hyperuniformity  on local structure has remained largely unexplored; establishing this connection can help uncover long-range interaction mechanisms and detect hyperuniform traits in finite-size systems.
Here, we study the topological properties of hyperuniform point clouds by characterizing their persistent homology and the statistics of 
local graph neighborhoods. We find that varying the structure factor results in configurations with systematically different topological properties. Moreover, these topological properties are conserved for subsets of hyperuniform point clouds, establishing a connection between finite-sized systems and idealized reference arrangements. Comparing distributions of local topological neighborhoods reveals that the hyperuniform arrangements lie along a primarily one-dimensional manifold reflecting an order-to-disorder transition via hyperuniform configurations. The results presented here complement existing characterizations of hyperuniform phases of matter, and they show how local topological features can be used to detect hyperuniformity in size-limited  simulations and experiments.
\end{abstract}

\maketitle

\section{Introduction}

Hyperuniform (HU) systems are characterized by vanishing density fluctuations at large length scales \cite{TorquatoPRE2003}. This property is trivially present in ordered systems, whereas for disordered systems it realizes a distinguishable state of matter \cite{torquato2018hyperuniform} lying between (quasi)-crystalline and amorphous or liquid phases: HU systems feature long-range order while being statistically isotropic with no Bragg peaks. These systems can exist either as equilibrium or quenched nonequilibrium phases and exhibit a range of peculiar electronic, photonic, or other transport properties~\cite{torquato2018hyperuniform,yu2021engineered}.
HU arrangements emerge in diffusive systems \cite{JackPRL2015,SkolnickActaMater2023}, emulsions \cite{WeijsPRL2015}, amorphous materials \cite{ZhengSCIADV2020}, nanostructure self-assembly \cite{salvalaglio2020hyperuniform}, supercooled liquid and glasses \cite{ZhangSciRep2016,Wang2021,Mitra_2021}, vortexes in superconductors \cite{LeThien2017,Llorens2020,SanchezPRB2023}, avian photoreceptors \cite{JiaoPRE2014}, swimmers \cite{huang2021circular}, and cosmology \cite{Gabrielli2002,GabrielliPRD2003} (for which it was first termed \textit{superhomogeneity}).
Systems engineered to be HU have found many applications, including polarization sensitivity \cite{gerasimenko2019quantum}, lasing \cite{degl2016hyperuniform,froufe2017band}, coatings with unusual anti-reflective properties or appearance \cite{chehadi2021scalable,vynck2022visual} and
full photonic band gaps for light propagation 
\cite{gerasimenko2019quantum,granchiPRB2023,Vynck2023} within optics, as well as topologically protected electronic states \cite{mitchell2018amorphous} and  mechanical systems \cite{XuPRE2017}. HU patterns also attract significant attention from statistics and probability theory 
\cite{ghosh2017number,KlattNatComm2019,klatt_last_2022}. 

Proper analysis of HU patterns is necessary for thorough explorations and detection of their distinctive properties. Although it has become straightforward to generate HU point patterns~\cite{Uche2004, Uche2006, florescu2009designer,Morse2023}, the identification and characterization of HU patterns in both experimental systems and numerical simulation frameworks remains challenging: HU characteristics are only rigorously defined for infinitely large systems \cite{torquato2018hyperuniform},  whereas any experimental or simulated data set is finite in size, and often only contains hundreds of points, see for instance \cite{LeThien2017,Llorens2020,Wang2021,Mitra_2021}. Consequently, their characterization is based on empirical diagnostics, essentially extrapolating measurements from finite length scales to infinity. Recent analyses have improved this statistical estimation ~\cite{hawat2023estimating}. However, an ideal HU character may be lost due to the presence of defects or perturbations even in strongly correlated systems \cite{chen2021topological,Tsabedze_2022,PuigDEF}, although the corresponding pattern should retain some essential properties of the ideal HU arrangements.
In brief, HU patterns exhibit local structures that we would like to characterize and explore in finite systems, aside from infinite wavelength measurements. Here, we pursue a topological approach, which by its nature, is robust to small changes in the system, providing an orthogonal way to characterize and explore HU structure without estimating global properties directly. 


Specifically, we aim to characterize HU point configurations by applying methods from topological data analysis (TDA). The TDA  framework employs techniques from algebraic topology to extract and analyze structures from complex datasets~\cite{Wasserman2018}. A method of particular interest to our study  is \textit{persistent homology}
\cite{edelsbrunner2002topological, edelsbrunner2008persistent, edelsbrunner2022computational}, which enables the robust and compact characterization of topological features across multiple scales for point clouds and other discrete data~\cite{ROBINS2016}.  
Practical computational methods have been developed to implement persistent homology calculations, and these tools have been applied across a range of disciplines~\cite{otter2017roadmap, aktas2019persistence}. 
Furthermore, we will also investigate the statistics of local topological neighborhoods in HU point patterns derived from the Delaunay tessellation~\cite{VoronoiBook, LazarRycroft}, using a recently introduced framework~\cite{Skinner2021, skinner2022topological}. The Delaunay tessellation characterizes the topological neighborhood structure, and whilst it will be different for each specific realization, we can robustly quantify its statistical properties, essentially counting how frequently each local topological neighborhood motif occurs in a point pattern. As these neighborhoods only change through discrete topological transitions, counting the number of transitions needed to transform one system into another gives a measure of the distance between HU systems 
~\cite{Skinner2021, skinner2022topological}. Altogether, our results demonstrate that  topological approaches can provide robust characterizations of HU configurations, even for finite systems.

The paper is organized as follows. In Sec.~\ref{sec:HU}, we briefly review the basic notions of HU systems and describe the generation and parametrization of point clouds.
In Sec.~\ref{sec:PH} we study the persistent homology of these patterns, quantifying differences and similarities between patterns. We also investigate how the underlying HU character can be detected in subsets of generated HU configurations. Sec.~\ref{sec:neighborhoods} analyzes the statistics of local graph neighborhoods, which enables comparisons between HU configurations. We find that many different HU configurations lie on a primarily one-dimensional manifold reflecting an order-to-disorder transition. The main conclusions are summarized in Sec.~\ref{sec:conclusions}.

\section{Hyperuniformity and generation of point clouds} 
\label{sec:HU}

An arrangement of points in a $d$-dimensional Euclidean space is said to be HU if the variance $\sigma_N^2(R)$ of the number of points in a $d$-dimensional spherical observation window with radius $R$ scales slower than the volume of this  window \cite{TorquatoPRE2003}, 
\begin{equation}\label{eq:HUdef1}
\lim_{R\rightarrow\infty} \frac{\sigma^2_N(R)}{R^d}=0.
\end{equation}
Rather than work with the variance directly, it is equivalent, and often easier, to define hyperuniformity in terms of the structure factor for an $N$ particle system \cite{torquato2018hyperuniform}
\begin{equation}\label{eq:sf}
    S(\mathbf{k},\rset)=\frac{1}{N}\bigg|\sum_{j=1}^N\exp({\rm i}\mathbf{k}\cdot\mathbf{r}_j)\bigg|^2, 
\end{equation}
with $\mathbf{r}_j$ the position of particle $j$, and $\mathbf{k}$ a wave vector. 
Formally, we only define $S(\mathbf{k},\rset)$ at a finite number of $\mathbf{k}=(k_1,\dots,k_d)$ vectors with components $k_i=2\pi n/L_i$, confining our particles in a $d$-dimensional box with side lengths $L_i$.
In the $N\to\infty$ limit, $S(\mathbf{k},\rset)$, simply referred to as $S(\mathbf{k}$) hereafter, becomes a continuous function of $\mathbf{k}$, aside from a formal singularity at $\mathbf{k}=\mathbf{0}$ (note
$S(\mathbf{0})=N$)~\cite{torquato2018hyperuniform}. 
Having taken this limit, a definition of a HU system equivalent to \eqref{eq:HUdef1} is \cite{torquato2018hyperuniform}
\begin{equation}\label{eq:HUdef2}
    \lim_{|\mathbf{k}|\rightarrow 0}S(\mathbf{k})=0.
\end{equation}
Both conditions \eqref{eq:HUdef1} and \eqref{eq:HUdef2} can be realized with different scalings of $\sigma_N^2(R)$ and $S(\mathbf{k})$. For systems characterized by a power law  $S(\mathbf{k})=|\mathbf{k}|^\alpha$ with $\alpha>0$ for $|\mathbf{k}|\rightarrow 0$, three characteristic scaling behaviors are realized, namely
\begin{equation}\label{eq:classes}
\sigma_N^2(R)\sim 
\begin{cases}
    R^{d-1} & \alpha>1 \\
    R^{d-1}\ln(R) & \alpha=1 \\
    R^{d-\alpha} & 0 < \alpha < 1 
\end{cases},
\end{equation}
with these three conditions referred to as HU classes I, II, and III, respectively~\cite{torquato2018hyperuniform}.

        
\begin{figure}[t]
    \centering
    \iflong
    \includegraphics[width=\linewidth]{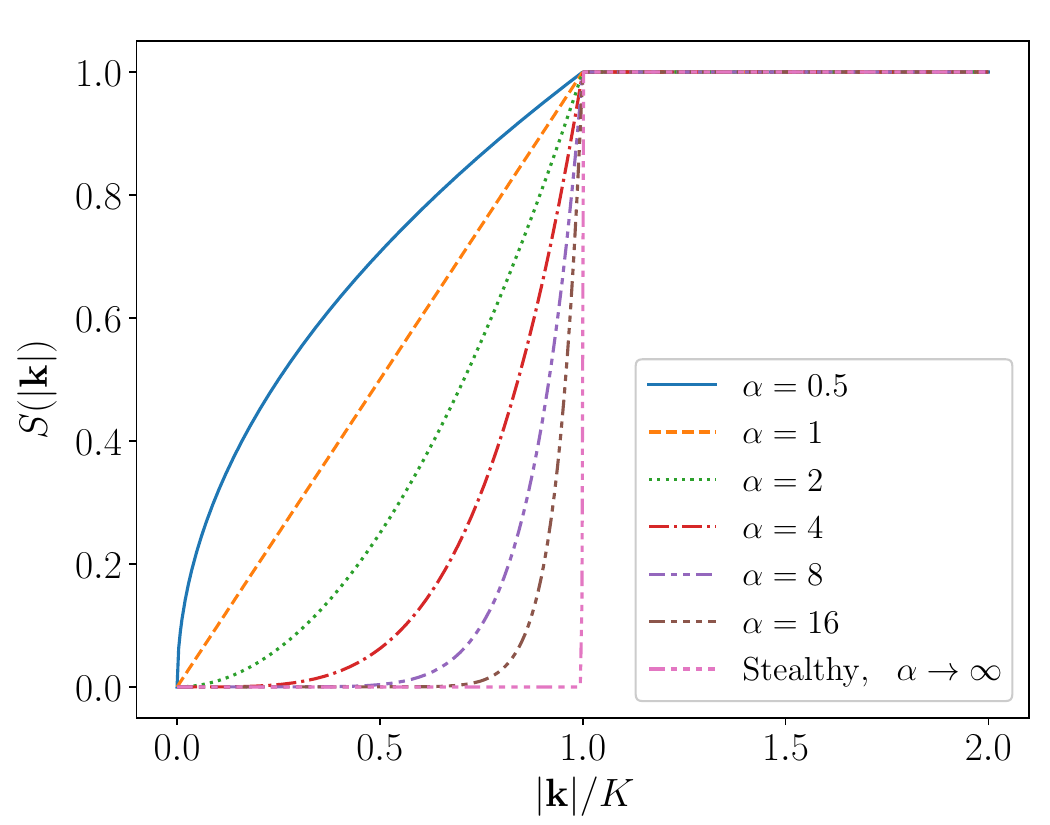}
    \fi
    \caption{The parameterized structure factor $S_0(|\mathbf{k}|)$ from Eq.~\eqref{eq:Sexample} for different $\alpha$ and $H=0$, showing hyperuniformity ($S(|\mathbf{k}|) \to 0$ as $\mathbf{k}\to 0$).}
    \label{fig:figure1}
\end{figure}

\begin{figure*}
    \centering
    \iflong
    \includegraphics[width=\linewidth]{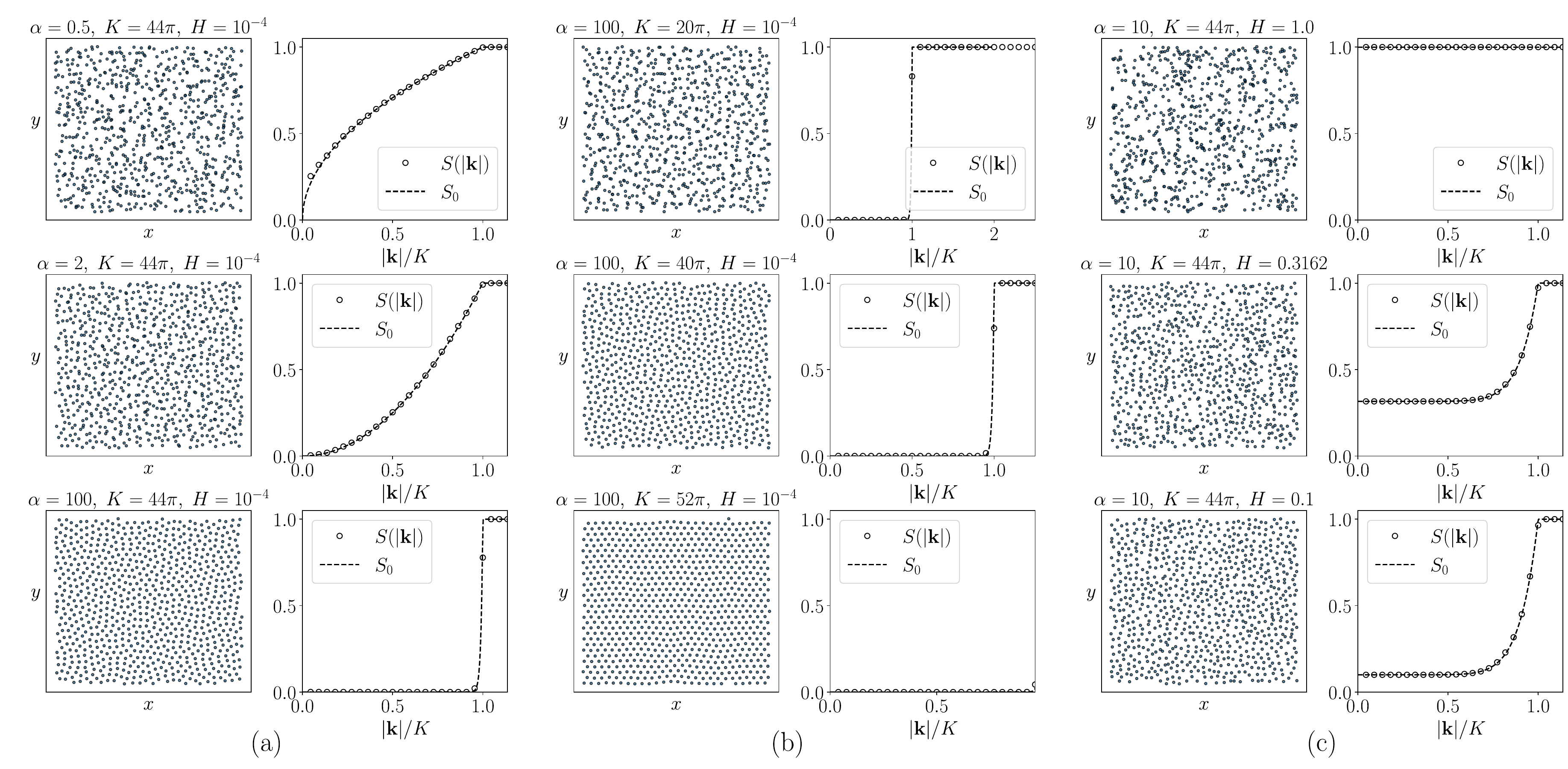}
\fi
    \caption{Examples of HU and non-HU point patterns ($N=780$) generated as described in Sec.~\ref{sec:HU}, shown together with the radial average of the structure factor $\Sk$ and the target structure factor, $S_0$. (a) $K=44\pi$, $H=10^{-4}$ with $\alpha=0.5$, $2$ and $100$ from top to bottom. For $\alpha=100$ an effective stealthy HU point cloud is obtained. (b) $\alpha=100$, $H=10^{-4}$ with $K=20\pi$, $40\pi$ and $52\pi$ from top to bottom. These are all stealthy HU point clouds and the order-disorder transition is expected for $K\approx 45\pi$. Note that for $K=52\pi$ a slightly perturbed triangular lattice is obtained.
    (c) $\alpha=100$, $K=40\pi$, with $H=1.0$, $0.3162$ and $0.1$ from top to bottom. $\Sk$ is shown for the same range of $|\mathbf{k}|$ ($|\mathbf{k}|\in[0,52\pi]$) in all the plots.}
    \label{fig:figure2}
\end{figure*}

Point clouds with a prescribed structure factor, $S_0(\mathbf{k})$, can be generated by solving an optimization problem~\cite{Uche2004, Uche2006}. In short, from an initial $N$-point arrangement 
with position of points $\rset$ given,  minimize an objective function, 
\begin{equation}
    F(\rset)=\sum_{\mathbf{k}}|S(\mathbf{k})-S_0(\mathbf{k})|^2,
\end{equation}
resulting in a least-square problem, where the sum over $\textbf{k}$ is taken over wave vectors appropriate for the finite-size bounding box. 
The minimization can be performed with standard algorithms like the conjugate-gradient, Broyden–Fletcher–Goldfarb–Shanno, or dogleg algorithms \cite{Uche2004,Uche2006}. 

We consider here a tunable, prototypical form for a power-law behavior of $S_0$ as follows (see also Fig.~\ref{fig:figure1})
\begin{equation}\label{eq:Sexample}
   S_0(\mathbf{k}) =  \begin{cases}
    D(1-H) |\mathbf{k}|^\alpha + H & \ |\mathbf{k}| < K, \\
    1  & \text{otherwise},
    \end{cases}
\end{equation}
with $D=K^{-\alpha}$ and $H$, $K$, and $\alpha$ three parameters controlling specific pattern features as detailed below.

The scaling exponent $\alpha$ controls the class of HU as defined in Eq.~\eqref{eq:classes}, and thus the overall shape of the structure factor, Fig.~\ref{fig:figure1}.
The limit of large $\alpha$ gives a finite-size region with $S(\mathbf{k})=H$, a feature referred to as \textit{stealth hyperuniformity} for $H \approx 0$ 
(see Fig.~\ref{fig:figure2}a, with a stealthy HU point cloud effectively obtained for $\alpha=100$ and $H=10^{-4}$).

Loosely speaking, $K$ sets the length scale, $2\pi/K$, above which correlations are enforced in the system. A small $K$ only suppresses fluctuations on the largest scales, allowing for more locally disordered arrangements whereas a large $K$ constrains arrangements at the size of a typical neighborhood. For stealthy HU configurations ($S(\mathbf{k})\sim 0$ for $0 <|\mathbf{k}|\leq K$), a known transition from disordered to ordered systems, signaled by the emergence of Bragg peaks in $S(\mathbf{k})$, is obtained at \cite{torquato2018hyperuniform,torquatoPRX2015}
\begin{equation}
    K = c_d \rho^{1/d},
\end{equation}
where $\rho=N/V$ is the density, $d$ the system dimensionality, and $c_d$ a (known) constant. In 2D, for $V=1$, stealthy HU disordered configuration are obtained for $K<\sqrt{8\pi N}$ (see also Fig.~\ref{fig:figure2}b, for which $\sqrt{8\pi N}\approx 45\pi$). 

The parameterization in  \eqref{eq:Sexample} is chosen so that $H = \widetilde{H}$, where
\begin{equation}
    \widetilde{H}[S]=\lim_{|\mathbf{k}|\rightarrow 0}\frac{\Sk}{S(k_{\rm max})},
\end{equation}
with $k_{\rm max}$ the characteristic wavenumber at which $S(\mathbf{k})$ assumes its maximum value. Formally, $\widetilde{H}=0$ for HU patterns (Eq.~\eqref{eq:HUdef2}),  although for finite-size systems $\widetilde{H}$ cannot be directly evaluated and must be extrapolated, as $|\mathbf{k}|\to 0$ implies a system with infinite size~\cite{torquato2018hyperuniform,salvalaglio2020hyperuniform}. Moreover, this quantity may deviate from zero due to the presence of defects or data artifacts (e.g., in experimental images). A commonly accepted convention is to say a point pattern is \textit{nearly} HU for $\widetilde{H}\leq 10^{-2}$ and \textit{effectively} HU for $\widetilde{H}\leq 10^{-4}$ \cite{Torquato2006,Kim2018}. By tuning the parameter $H$, we can study the convergence to HU, and how that alters the topology of the patterns (see also Fig.~\ref{fig:figure2}c).


In the following, we will consider $d = 2$ space dimensions and generate point patterns for $N=780$ in a square 2D domain with unitary sides, as illustrated in Fig.~\ref{fig:figure2}. $N$ is chosen such that the points can be arranged almost as an ideal triangular arrangement (closest packing in 2D). Such a lattice can be obtained by arranging $2q$ rows of $p$ particles with $p/q$ a close approximation of $\sqrt{3}$. A set of possible $p$,$q$ choices is reported in \cite{Uche2004}. We obtain patterns by varying parameters in the following ranges
\begin{equation}\label{eq:parameters}
\begin{split}
H& \in [10^{-4},1], \\
\alpha& \in [0.5,100], \\
K &\in [20\pi,56\pi].
\end{split}
\end{equation}
We generated 880 patterns by varying the parameters in the range shown in Eq.~\eqref{eq:parameters}. All combinations start from a periodic point pattern with relatively small random noise before minimizing $F$~\cite{Uche2004,Uche2006}, with ten different random seeds, and hence ten final patterns for each parameter combination.

Examples of generated patterns, with varying $K$, $\alpha$, and $H$ parameter values, are shown in Fig.~\ref{fig:figure2} together with the radial average of the structure factor, $\Sk$, and the target structure factor $S_0$.  Varying $\alpha$ for a small $H$ (effective HU) and a fixed $K$ close to the nominal order-disorder transition for stealthy HU patterns explores the three different HU classes from Eq.~\eqref{eq:classes}, where increasing $\alpha$ leads to more and more uniformly distributed points (Fig.~\ref{fig:figure2}, first column). A qualitatively similar effect is observed by varying $K$ (Fig.~\ref{fig:figure2}, second column). For small $K$, even a stealthy HU configuration leads to small-scale agglomerates because no correlation is imposed for length scales smaller than $2\pi/K$. Approaching the order-disorder transition leads to a more uniform arrangement, while above the critical point, a (slightly perturbed) triangular arrangement is obtained. The effect of $H$ is illustrated first in the limiting case of $H=1$, resulting in a random point pattern, and for two other values (Fig.~\ref{fig:figure2}, third column) showing the impact of lowering the small wave vector limit of $S$ (still far from a nearly HU pattern). Although the trends appear clear, it remains challenging to quantify differences and similarities between these patterns, particularly when far away from limiting cases (random or ordered). In the remainder, we show how one can overcome this problem by studying and characterizing the topology of these point patterns.

\section{Persistent Homology}
\label{sec:PH}

\begin{figure}[t]
    \centering
    \iflong
    \includegraphics[width=\linewidth]{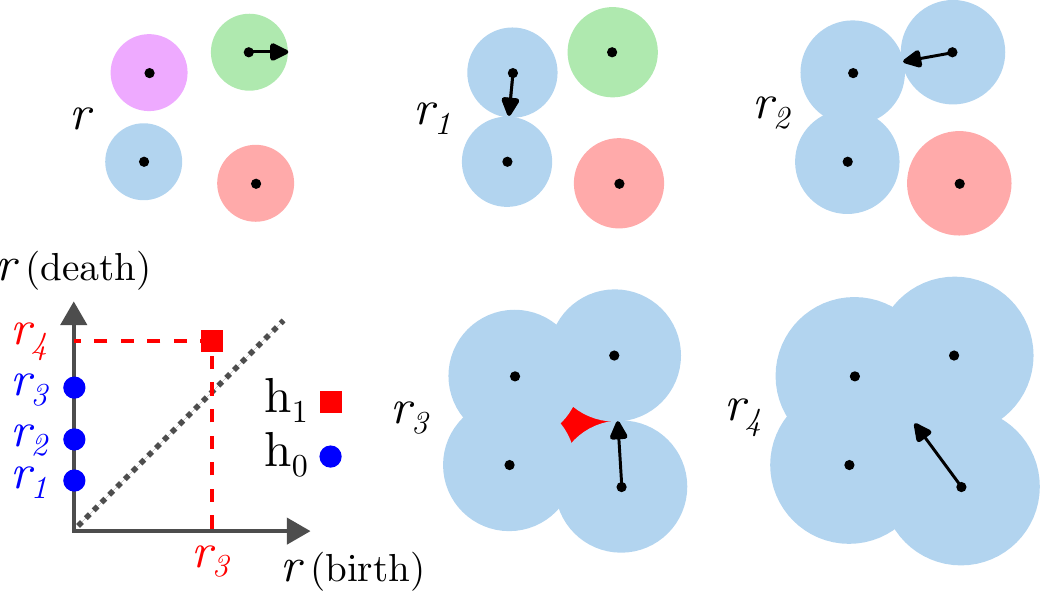}
    \fi
 \caption{Illustration of the persistence diagram (bottom left) for a four-point arrangement (top left), and a family of nested Vietoris-Rips complexes. We parameterize the complex by $r$, where $r$ is the radius (\textit{time}) of circles centered on the points. Shown are radii,  $r_i$, at which features are added to the persistence diagram. Blue points on the persistence diagram represent the birth/death of $h_0$, or connected components, including initial isolated circles which all have a birth time of 0, and death times of $r_{1,2,3}$. The red square represents $h_1$, the hole (or closed loop) within the domains, forming at $r_3$ and vanishing at $r_4$.}
    \label{fig:figureIllPD}
\end{figure}

Persistent homology \cite{edelsbrunner2002topological,edelsbrunner2008persistent,edelsbrunner2022computational} uses ideas from algebraic topology to understand the ``shape'' of a data set, building on how homology characterizes geometric objects through connected components, loops, and voids (Betti numbers). 
Starting from a point set $P$, persistent homology constructs a family of simplicial complexes (collection of points, lines, triangles, tetrahedrons, and higher dimensional analogs), $K_P^r$, parameterized by $r\geq 0$, such that $K_P^l \subseteq K_P^m$ for $l<m$. This nested family of simplicial complexes is called a \textit{filtration}. Persistent homology takes the appearance (\textit{birth}) and disappearance (\textit{death}) of homological features of this filtration, as $r$ is varied, as the essential topological information that characterizes the point set. 
A graphical representation for the births and deaths of topological features in the filtration is called the \textit{persistence diagram}~\cite{otter2017roadmap,aktas2019persistence,chazal2021introduction}, see Fig.~\ref{fig:figureIllPD}, which is discussed further below. 

\begin{figure*}
    \centering
    \iflong
    \includegraphics[width=\linewidth]{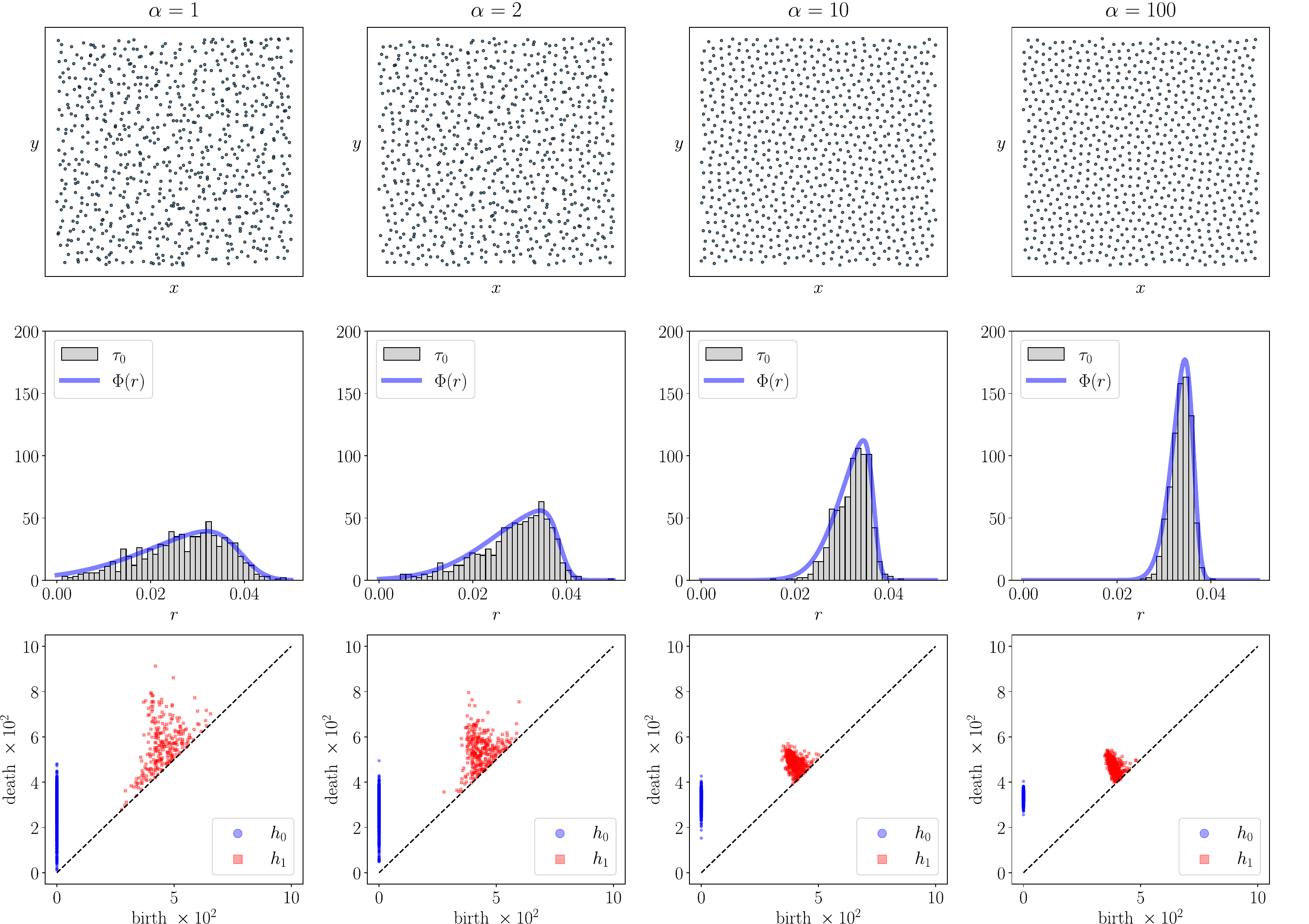}
    \fi
 \caption{Persistent homology analysis for representative patterns featuring different $\alpha$ with $K=44\pi$ and $H=10^{-4}$. The first row shows four-point patterns. The second row illustrates the death times distribution $\tau_0$ for connected components $(h_0)$ through a histogram and a fit with the skew-normal distribution $\Phi$. The third row shows persistence diagrams for $h_0$ (connected components) and $h_1$ (holes). Additional results obtained by varying $K$ and $H$ are reported in the Supplemental Material.}
    \label{fig:figurePD}
\end{figure*}
For our choice of filtration, we use the widely adopted Vietoris-Rips complex \cite{edelsbrunner2022computational}. In short, consider placing a circle of radius $r$, which we use as the filtration parameter, around each point in the data set. If two circles overlap, the line connecting the two corresponding points is included in the simplicial complex. In other words, two separated connected components, the points, become one connected component. If three circles all pairwise overlap, then the triangle defined by the three corresponding points is included in the simplicial complex, and so on, see Fig.~\ref{fig:figureIllPD}. We increase the radius $r$, which we hereafter refer to as \textit{time}~\cite{otter2017roadmap,aktas2019persistence}, from zero to infinity, and consider how the resulting topology of the simplicial complex changes.  This procedure echoes the concept of diffusion spreadability in a two-phase medium, namely the time-dependent mass transfer from one phase (say, circles) into a second phase (space surrounding the circles) which, interestingly, shows peculiar behaviors for disordered HU configurations~\cite{torquato2021diffusion}. A topological feature, such as a hole, is created at some time and disappears at a later time, Fig.~\ref{fig:figureIllPD}. This can be represented as a point on the persistence diagram, with the birth and death times of this feature as the $x$ and $y$ coordinates, respectively. The persistence of a feature is the difference between its death time and birth time, indicating how long the feature exists as the parameter $r$ changes. For our analysis, we will look at the birth/death of connected components, $h_0$, and of holes or loops, $h_1$ (see Fig.~\ref{fig:figureIllPD}). In general, $h_k$ generalizes the concept of holes through homology: $h_k$ quantifies $k$-dimensional holes \cite{edelsbrunner2022computational}, with zero-dimensional holes then corresponding to connected components and one-dimensional holes corresponding to closed loops or ``holes'' in an informal sense. Persistent diagrams are computed using the python library \texttt{ripser} \cite{ctralie2018ripser,Bauer2021Ripser}.

\begin{figure*}
    \centering
    \iflong
\includegraphics[width=\linewidth]{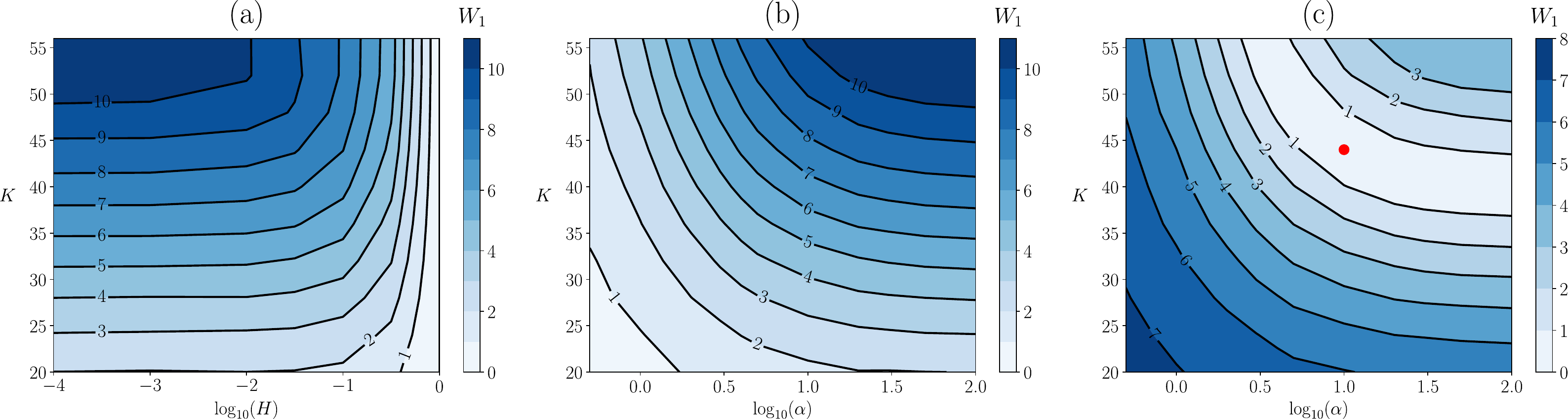}
    \fi
    \caption{Illustration of distances $W_1(h_0^{\rm (I)},h_0^{\rm (II)}$) with $\rm (I)$ and $\rm (II)$ labelling the corresponding point patterns from which persistence diagrams for $h_0$ (connected components) are computed. In panels (a) and (b), ${\rm (I)}$ is a random arrangement and ${\rm (II)}$ is varied with (a) different $H$ and $K$ with $\alpha=100$ (corresponding to the stealthy HU  regime for sufficiently small $H$), and (b) different $\alpha$ and $K$ with $H=10^{-4}$. In panel (c), we show the distance from a selected HU pattern: ${\rm (I)}$ is an arrangement featuring $K=44\pi$, $H=10^{-4}$, $\alpha=10$ and ${\rm (II)}$ patterns featuring different $\alpha$ and $K$ with $H=10^{-4}$. The parameter corresponding to the pattern ${\rm (I)}$ is marked as a red dot.}
    \label{fig:figureWD}
\end{figure*}

\begin{figure*}
    \centering
    \iflong
     \includegraphics[width=\linewidth]{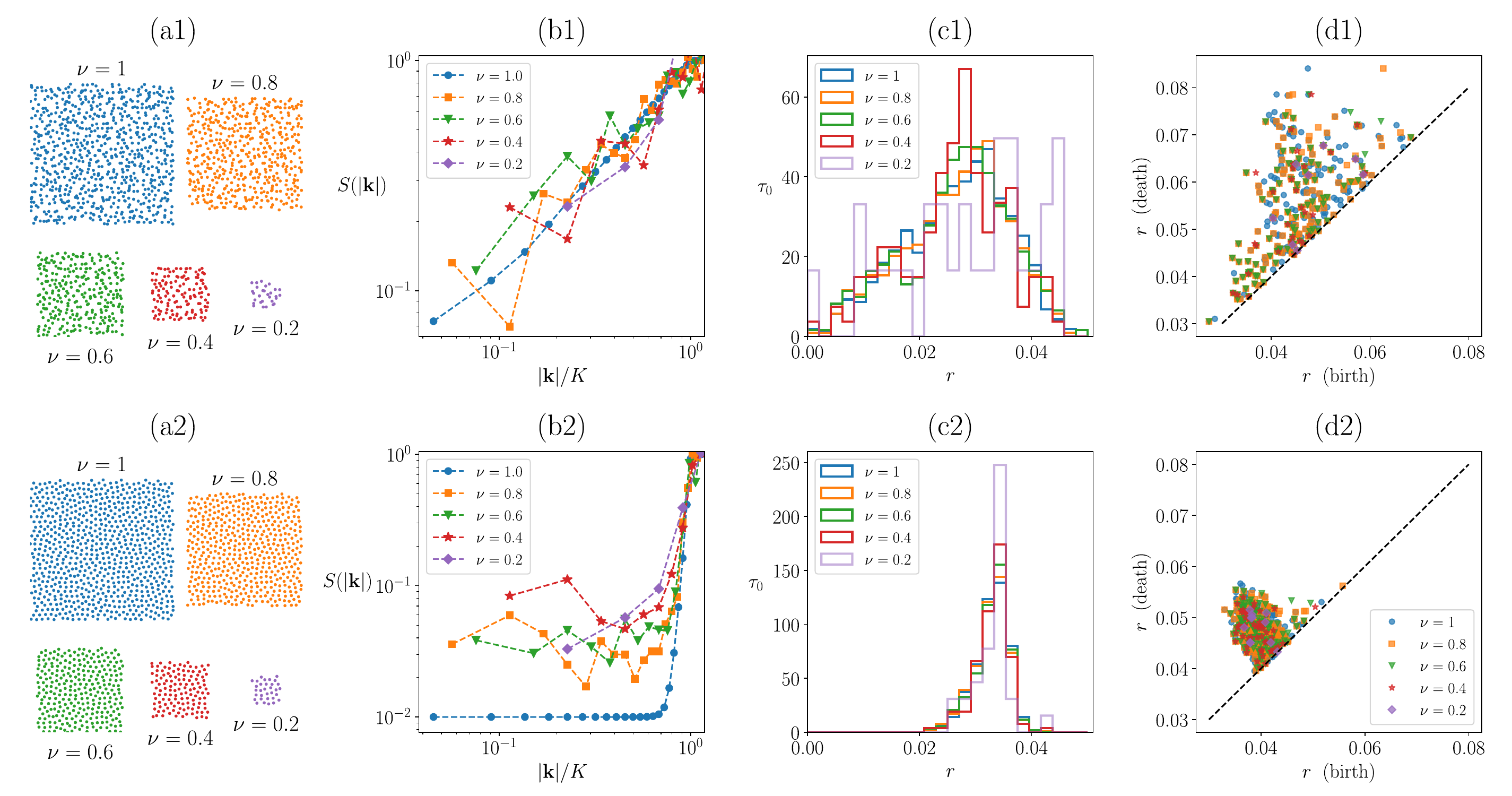}
    \fi
    \caption{Topological properties of an ideal HU point pattern are conserved in subsets of that point pattern. (a1)-(a2) Different sets obtained by selecting a $\nu \times \nu$ region of a pattern generated as in the previous section ($\nu=1$) with $K=44\pi$, $H=10^{-2}$, (a1) $\alpha=1$ and (a2) $\alpha=20$. For these patterns we show: (b1)-(b2) S$(|\mathbf{k}|)$, (c1)-(c2) histograms $\tau_0$ (d1)-(d2) $h_1$ persistence diagrams.}
    \label{fig:figureSZ}
\end{figure*}  

Figure~\ref{fig:figurePD} shows persistent homology analysis applied to HU patterns, here obtained by varying $\alpha$ whilst keeping the other parameters constant. Starting from the HU point patterns (first row), we can compute a histogram of death times $\tau_0$ for connected components (second row), and persistence diagrams of connected components, $h_0$, and holes, $h_1$ (third row). The same analysis for patterns obtained by varying $K$ and $H$ is reported in the Supplemental Material. Clear trends emerge, with a narrowing of the distributions for more pronounced HU characters (larger $\alpha$ and $K$, as well as smaller $H$). Note that significant variations in the topological features are also observed for patterns that closely resemble each other; cf. $\alpha=10$ and $\alpha=100$ in Fig.~\ref{fig:figurePD}. We find that a skew-normal distribution fits histograms of the death times $\tau_0$ well, see Fig.~\ref{fig:figurePD} (second row). Clear trends in the parameters of the skew-normal also emerge when varying $\alpha$, $K$, and $H$, see the Supplemental Material for details.

To quantify differences between patterns from the topological features obtained above, we use the Wasserstein distance between two persistence diagrams, specifically the distributions of death/birth times for $h_0$ (connected components) and $h_1$ (holes). The $p$-th Wasserstein distance between two persistence diagrams $A$ and $B$ (including the diagonal) is defined as \cite{aktas2019persistence}
\begin{equation}\label{eq:WD}
W_p(A,B)=\inf_\gamma \bigg( \sum_{\mathbf{x} \in A} (||\mathbf{x}-\gamma(\mathbf{x})||_q)^p\bigg)^{1/p} ,
\end{equation}
with $\gamma(\mathbf{x})$ ranging over all bijections from $A$ to $B$ and $||\mathbf{a}-\mathbf{b}||_q=(\sum_{i=1}^2 |a_i-b_i|^q )^{1/q}$ (as $\mathbf{a},\mathbf{b}\in \mathbb{R}^2$). Here, $q$ determines the cost of transporting $\mathbf{x}$ to $\gamma(\mathbf{x})$; for example, if $q = 1$, the cost is the $L^1$ (Manhattan) distance, and if $q = 2$ the cost is the Euclidean distance. Throughout, we set $p=q$. Overall, $W_p$ quantifies the similarities between two sets of points by computing the minimal cost to transform one set of points into another.  Calculations involving Wasserstein distances between persistence diagrams used the
\texttt{GUDHI} libraries \cite{gudhi:urm}. Other methods to compare persistence diagrams or related barcodes are summarized in \cite{ali2022survey}.

\begin{figure*}[t]
    \centering
    \iflong
\includegraphics[width=\linewidth]{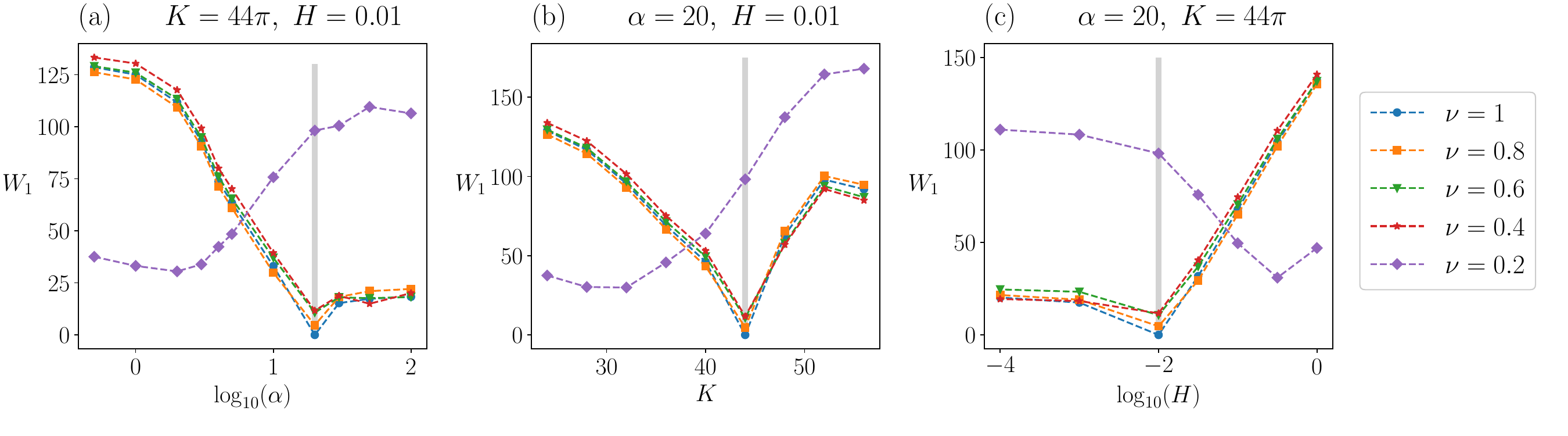}
\fi
\caption{Topological properties of finite subsets of a HU point pattern are closest to reference patterns generated with the same parameters. We compute $W_1(\tau_0^{\rm (I)},\tau_0^{\rm (II)})$ with (I) the patterns in Fig.~\ref{fig:figureSZ}(a2) and (II) reference arrangements generated with different $\Sk$ parameters, averaged over ten numerical realizations. Different panels show variation of one parameter, namely (a) $\alpha$, (b) $K$, (c) $H$, while others are held constant for values specified in the figure. Vertical grey lines show the actual value used to generate the subset data Fig.~\ref{fig:figureSZ}(a2).}
    \label{fig:figureSZ2}
\end{figure*}

Figure~\ref{fig:figureWD} illustrates $W_1$ between $h_0$ (connected components) persistence diagrams of a random pattern and patterns obtained by varying (a) $H$ and $K$ with $\alpha=100$ and (b) $\alpha$ and $K$ with $H=10^{-4}$, both featuring stealthy HU configurations for sufficiently small $H$ and large $\alpha$. The qualitative trends emerging in Fig.~\ref{fig:figure2} are quantified by $W_1$, showing that, as $K$ and $\alpha$ increase, the pattern becomes less similar to a random pattern, whereas when $H$ increases the pattern looks more like a random pattern, Fig.~\ref{fig:figureWD}. 
From these results, we see that the topological properties of the ideal limit $H \sim 0$ are effectively obtained for $H<10^{-3}$ (see, Fig.~\ref{fig:figureWD}a). No further effects are expected by lowering $H$ below $10^{-4}$ for the domain size and number of points considered here. This is a useful estimation as $H$ quantifies HU character in realistic settings: in a system with hundreds of points, we find that $H<10^{-3}$ leads to topologies sufficiently similar to the ones of ideal HU point clouds. 
Similar results are obtained with $W_2$ between persistence diagrams for $h_0$; see Supplemental Material. Distances (both $W_1$ and $W_2$) between persistence diagrams for $h_1$ (holes) follow similar trends for a broad range of $K$ and $\alpha$ values. However, distances from random patterns slightly decrease for increasingly large values of $K$ and $\alpha$ (ordered systems). Thus, comparing $h_1$ is somewhat less effective at characterizing topological differences than comparing $h_0$.

Alternatively, we can measure the distance of persistent diagrams to a reference one obtained for a HU arrangement. This is illustrated in Fig.~\ref{fig:figureWD}c, which shows $W_1$ between $h_0$ (connected components) persistence diagrams of a reference configuration corresponding to a pattern obtained with $K=44\pi$, $\alpha=10$ and $H=10^{-4}$ (red dot in Fig.~\ref{fig:figureWD}c), and ones obtained by varying $\alpha$ and $K$. $W_1$ generally increases when moving towards random (small $K$ and $\alpha$) or ordered (large $K$ and $\alpha$) arrangements. Interestingly, a vanishing distance is obtained for the reference configuration (at the red dot) as well as along a curve in the $K$-$\alpha$ parameter space passing through this point. Analogous topological properties are thus obtained with different parameter combinations for $S(\mathbf{k})$. Said otherwise, there exists a manifold in the parameter space where  patterns possess analogous topological properties. The same conclusion can be drawn from $W_2$ and $h_1$ (holes) persistent diagrams; see Supplemental Information. This concept will be explored further in section~\ref{sec:neighborhoods}.


We now investigate how the structure factor and topological properties are affected by finite-sized data. We consider two specific arrangements of points generated as above with $H=10^{-2}$, a nearly-HU pattern, and $K=44\pi$, $\alpha=1$ as well as $K=44\pi$, $\alpha=20$, resulting in visually distinct point patterns, Fig.~\ref{fig:figureSZ}a1-a2. We select points within $\nu \times \nu$ squares of different sizes with $\nu \leq 1$ ($\nu=1$ is the original arrangement), see Fig.~\ref{fig:figureSZ}a. Whilst the structure factor $S(\mathbf{k})$ matches the desired $S_0$ at $\nu=1$ by construction, for subsets $\nu<1$ it deviates from that of the full data ($\nu=1$), Fig.~\ref{fig:figureSZ}b1-b2. These subsets, which are not periodic and mimic realistic data where only a portion of the system can be analyzed, highlight the issue with characterizing HU arrangements by an inferred $S(\mathbf{k})$. Instead, we leverage topological features to assess how similar a finite-size pattern is to an ideally generated HU arrangement. For instance, one may look at the normalized histograms $\tau_0$ (death times of connected components), see Fig.~\ref{fig:figureSZ}c1-c2. These distributions have a similar shape for $\nu \geq 0.4$, pointing at some robust topological properties only deviating for small $\nu$ ($\nu=0.2$). Similarly, the persistence diagrams for $h_1$ (holes) appears conserved across $\nu$, Fig.~\ref{fig:figureSZ}d1-d2.

To quantitatively assess the extent to which  topological properties, here $\tau_0$, are conserved for finite-sized data, we computed the $W_1$ distance between the $\tau_0$ distribution of the five patterns in Fig.~\ref{fig:figureSZ}a2 and reference ideal patterns generated by varying separately $\alpha$, $K$, $H$, see Fig.~\ref{fig:figureSZ2}. The reference distribution was created using ten simulated repetitions for each ideal pattern. The finite-sized data was closest to the correct reference parameters in all cases for $\nu\geq 0.4$, Fig.~\ref{fig:figureSZ2}. The minimum distance obtained for the smallest subset ($\nu=0.2$) deviates significantly due to the extremely small number of points, reflecting a statistical lower bound on the size of the point set required to correctly identify its topological features. Consistent with observations reported above, using $W_2$ also finds that the finite-sized patterns are closest to the correct reference pattern for $\nu\geq0.4$.


We conclude that topological properties of point patterns set by an ideal $S(\mathbf{k})$ are still present in subsets of the same arrangements. In this case, we compared distributions of $\tau_0$ with a $W_1$ distance as an example, but other topological features or distances can be used. 

\section{Topological structure of neighborhoods}
\label{sec:neighborhoods}

\begin{figure*}
    \centering
    \iflong
    \includegraphics[width=\linewidth]{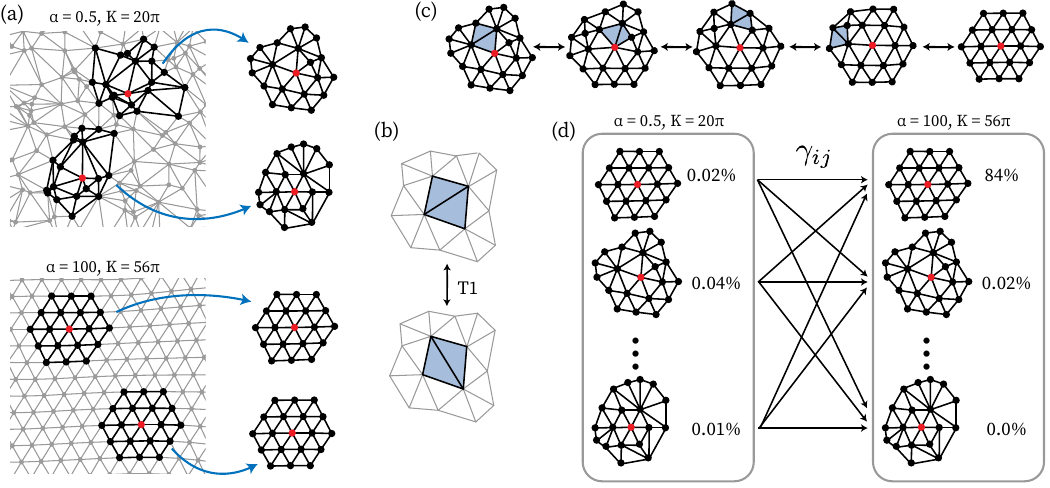}
    \fi
    \caption{Comparing two HU patterns based on the distribution of local topological networks or motifs. (a) Starting from a HU point pattern, shown for two parameter values ($\alpha =0.5$, $K=20\pi$, top and $\alpha = 100$, $K=56\pi$, bottom), we take the Delaunay tessellation. The local network of radius $r=2$ is the subgraph induced by neighbors and nearest neighbors around a central vertex; four examples are shown in black with a red central vertex. (b) The Delaunay tessellation is a topological object and only changes through discrete T1 transitions or flips. (c) Any motif can be transformed into any other motif through a finite sequence of flips. The motif on the left is transformed into the motif on the right with as few flips as possible. The triangles to be flipped as one moves to the right are shown in blue. (d) For a given HU point pattern, there is a probability distribution of local motifs, each occurring with some frequency; to compare two HU patterns, we compare these distributions. We do this by solving an optimal transport problem, finding a transport map $\gamma$ between two distributions, with the cost of transforming motif $i$ to $j$ set by the number of flips to transform one motif into the other. In effect, the distance represents the minimal number of flips needed to transform the left distribution into the right one.}
    \label{fig:TopoExplain}
\end{figure*}

Beyond the topological features of connected components and loops (persistent homology), we now examine the related~\cite{Bauer2016} topological structure of neighborhoods~\cite{Skinner2021, skinner2022topological, Lazar2015, LazarPRL}. Given a point pattern, we start from the corresponding Delaunay triangulation, dual of the Voronoi diagram, a topological object connecting neighboring points~\cite{VoronoiBook}. The Delaunay triangulation will be different for each realization of a point pattern. Still, by capturing its statistical properties, we can examine how the topology of neighborhoods systematically changes across different generating procedures. 

We use a recently introduced framework that statistically characterizes the local structure of the Delaunay triangulation and allows physically interpretable comparisons between different point patterns~\cite{Skinner2021, skinner2022topological}. Starting from the Delaunay triangulation, interpreted as a planar graph, for each vertex of this graph, we take the local \textit{neighborhood} of radius $r_g$; the induced subgraph formed by the set of all vertices which are at most $r_g$ edges away from the central vertex, Fig.~\ref{fig:TopoExplain}a. We use $r_g=2$  for computational and statistical purposes, which still results in tens of thousands of unique local neighborhoods~\cite{Skinner2021}. Two local neighborhoods are the same topological type, or \textit{motif}, if they are graph-isomorphic. We characterize the topo-statistical state of a point pattern $M$ as a probability distribution over the space of these motifs, where $P_M(i)$ is the probability of seeing motif $i$.

Delaunay triangulations are invariant under infinitesimal perturbation and change only through discrete topological transitions, Fig.~\ref{fig:TopoExplain}b. A natural distance between two motifs is then the minimum number of topological transitions required to transform one into the other, 
Fig.~\ref{fig:TopoExplain}c, a mathematically well-defined metric~\cite{lawson1972transforming}. Given two probability distributions $P_A$ and $P_B$ corresponding to different classes of point patterns, a natural distance between $A$ and $B$ is to measure how many transitions are required to transform distribution $P_A$ into $P_B$. Mathematically, this is a Wasserstein distance, known as a topological earth mover's (TEM) distance
\begin{equation}\label{eq:D}
    \text{TEM}(A,B)=\min_\gamma \sum_{ij}\gamma_{ij}d(i,j),
\end{equation}
with $d(i,j)$ the distance between motifs $i$ and $j$ and $\gamma_{ij}$ a transport map from $A$ to $B$ ($\gamma_{ij}\geq 0$, $\sum_j \gamma_{ij}= P_A(i)$, and $\sum_i \gamma_{ij}= P_B(j)$), which can be solved as a minimum cost flow problem on a network, Fig.~\ref{fig:TopoExplain}d. Interpreting each transition as having an energetic cost associated with it, the distance TEM$(A,B)$ can be interpreted as the average overall energetic cost to transform $A$ into $B$.

\begin{figure*}
    \centering
    \iflong
    \includegraphics{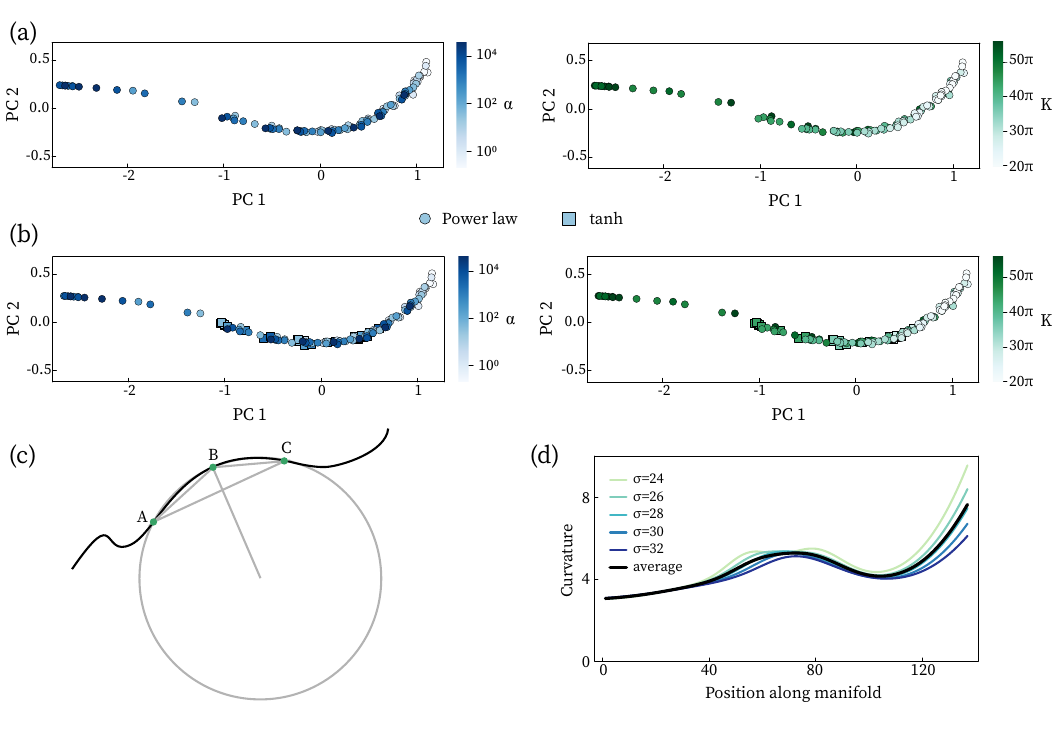}
    \fi
    \caption{Across a 2D region of parameter values, the resulting HU patterns lie on a nearly 1D manifold in a low dimensional embedding. (a) HU patterns were created for a range of $\alpha$ (left) and $K$ (right) parameter values for a power law $S_0$. The pairwise topological distance matrix was computed, which then was embedded into Euclidean space with MDS, resulting in an approximately 1D manifold when plotting the first two Principal Components (PCs); see Fig.~S8 in the Supplementary Material for PC3 and a residual variance analysis.  (b)~To test whether the manifold in (a) resulted from the specific power law form of $S_0$, we included patterns obtained with a tanh function, $S_0^T$. Including these additional patterns did not change the form of the embedding. (c) For three points, $A$, $B$, and $C$ (green) on an arbitrary curve (black), there is a unique circle that intersects those points (grey), with some radius $R$ and curvature $1/R$, which can be computed by knowing only the pairwise distances between $A$, $B$, and $C$. In the limit that $A,C \to B$, the limit of the curvature $1/R$ is the Menger curvature, a definition that can be extended to an arbitrary metric space. (d) Menger curvature as a function of position along the manifold in (b). Shown for different KDE widths, $\sigma$, as well as the average value.}
    \label{fig:WD}
\end{figure*}

Taking different point patterns generated by varying $\alpha$, $K$, and that are effectively HU ($H<10^{-3}$), we computed the pairwise distance between each of them, resulting in an $np \times np$ distance matrix $\mathcal{D}$, where $np$ is the overall number of unique parameters. To visualize this matrix, we find a low dimensional embedding with Multi-Dimensional Scaling (MDS) \cite{borg2005modern}, which finds the most faithful way to embed points in Euclidean space while matching the embedded Euclidean distances to the distance matrix $\mathcal{D}$. Interestingly, we find that in the embedded area, the data falls on an effectively one-dimensional manifold, Fig.~\ref{fig:WD}a, despite us varying two independent parameters (to quantify manifold dimensionality see Supplementary Material). We see a systematic trend on varying either $\alpha$ or $K$, where increasing either parameter moves us from the disordered side of the manifold (positive Principal Component 1) towards the ordered side of the manifold (negative Principal Component 1).

Generically, when varying $m$ parameters, one expects a $m$-dimensional data manifold, yet our effectively one-dimensional manifold means that topologically similar patterns can be obtained with different values in parameter space. This is consistent with our finding in Fig.~\ref{fig:figureWD}c, that there is a region of $\alpha-K$ parameter space at zero (or very small) $W_1$ distance from a given HU reference arrangement, suggesting that fewer than two parameters are needed to specify the topological features.

To test this property further, we 
consider another parametrization of $S_0$, which differs from the previous power-law definition, namely,
\begin{equation}
 S_0^{\rm T}=H+(1-H)\frac{1}{2}\bigg[1-\tanh\left(\frac{3(K^\prime-|\mathbf{k}|)}{\alpha^\prime\pi}\right)\bigg],
\end{equation}
with $K^\prime$ and $\alpha^\prime$ auxiliary parameters mimicking the role $K$ and $\alpha$ in Eq.~\eqref{eq:Sexample} (the same names without ``prime" will be used in the following).
Combining new point patterns with this modified parameterization, we calculate an expanded distance matrix and once again embed it in a low-dimensional space with MDS. The data still lies on a one-dimensional manifold, Fig.~\ref{fig:WD}b, showing that this is not just a property emerging from a particular form for $S_0$. Indeed, the space of topological neighborhoods for these HU point patterns appears to be one-dimensional, which is not a generic property of 2D point patterns~\cite{Skinner2021}.

Looking at the manifold, or curve, we see that it is not a straight line in the embedding. This means that the path taken along the curve is not optimal with respect to the number of topological transitions. Alternatively said, if one were to take an optimal path from, say $K=20\pi$, $\alpha = 0.5$ to $K=56\pi$, $\alpha=100$, then intermediate points on this path would not lie on the 1D manifold. The manifold appears to be maximally distorted for values of Principal Component 1 greater than zero. Still, one can not judge this solely by eye due to inherent distortions that occur when embedding a distance matrix in a low-dimensional Euclidean space.

Instead, we can turn to metric geometry to quantify how far from optimal the manifold path is without resorting to Euclidean embeddings. In plane geometry, three points, $A$, $B$, and $C$, uniquely define a circle. If those points lie on a curve, in the limit that $A, C \to B$, the circle converges, and the inverse radius of the limiting circle defines the curvature at $B$. The Menger curvature is the curvature defined by this procedure, but which calculates the radius of the circle using only the distances $AB$, $AC$, and $BC$, Fig.~\ref{fig:WD}c. This definition can then be extended to compute curvatures for any metric space, not just Euclidean ones, as it only uses distances. This allows us to quantify curvature along a 1D manifold in the topological space without ever embedding it into a lower-dimensional Euclidean space.

We can compute the Menger curvature at each point by parameterizing the 1D embedding manifold by a single parameter $t$, say the position along PC 1. As an aside, we could add new HU or non-HU patterns to the embedding and compute their value of $t$, finding where they lie along the manifold or whether they lie away from the manifold. Given the parameterization, $t$, we use a Gaussian kernel density estimator with width $\sigma$ to approximate a continuous probability distribution $p_t(i)$ for motif index $i$. The curvature can be computed from this by effectively regularizing the distance into a space with a unique geodesic~\cite{Skinner2021, skinner2022topological}. We find that the curvature is indeed smallest in the region corresponding to more ordered HU configurations and largest when approaching the region of disordered arrangements, Fig.~\ref{fig:WD}d. Interestingly, the curvature is always non-zero, reflecting that the path taken is not optimal. If our samples effectively span the space of HU patterns, then the curved nature of the manifold means that to get from one HU pattern to another, either you have to transition through a non-HU pattern or the path taken is not optimal with respect to neighborhood rearrangements.

\section{Conclusions}
\label{sec:conclusions}

In summary, by looking at the topological properties of HU configurations we can characterize and study their features in a way that is orthogonal to the structure factor. The persistence diagrams differentiate HU patterns across a wide range of parameters. Moreover,  distributions of topological properties are approximately conserved for finite-sized patterns. This enables the matching of these properties from a finite pattern, arising for instance from an experiment, to that of a reference ideal configuration.
This approach can complement the characterization of hyperuniformity with classical tools when directly estimating the structure factor is challenging. 
Additionally, we analyzed the topological structure of local neighborhoods, comparing point patterns through a topological earth mover's distance, which measures the number of rearrangements needed to go from one pattern to another. This also robustly identifies differences between topological properties in HU patterns.

Both distances between local graph-neighborhood motifs distributions and persistence diagrams show that different parameters used for the generation can lead to similar topological properties. Indeed, when varying two parameters in the structure factor, there was a one-dimensional curve along which ($h_0$ and $h_1$) persistence diagrams were equivalent under both the $W_1$ and $W_2$ distance (Fig.~\ref{fig:figureWD} and Supplemental Material). Similarly, in the space of distribution of local motifs, patterns generated by varying two parameters lie along a one-dimensional curve (Fig.~\ref{fig:WD}). The latter analysis leads to the same results after including patterns generated with different forms of the structure factor. On the one hand, this result points out that there is no unique correspondence between a structure factor parametrization and the resulting topological features, with important implications for inverse design. On the other hand, this demonstrates that one can control distinctive topological properties by varying \textit{one} effective parameter.

By analyzing the curvature of the 1D manifold in the space of distributions of local motifs \cite{Skinner2021}, we show that the variation of topological properties along the manifold does not correspond to the solution of a topological optimal transport problem. In other words, a system that transitions between different HU states may need to explore topologies that differ from characteristics of HU patterns or take a non-optimal path. Besides its theoretical relevance, this result could be used to interpret the emergence of HU states and the transition between HU and non-HU states in systems far from equilibrium \cite{huang2021circular, backofen2023nonequilibrium}.

The analysis provided here for prototypical HU point clouds can be straightforwardly extended to patterns featuring different kinds of correlated disorder, such as in disordered lattices~\cite{Keen2015, Llorens2020} or colloidal particles systems with clustering effects~\cite{Wang2021,wang2018magic,whitaker2019colloidal}. 
Moreover, an interesting question suggested by this work is whether finite, and ultimately experimental, patterns could be designed based on their topological properties rather than imposing directly the HU condition, which is not defined for finite systems. One could eventually explore their physical properties,  for instance their interaction with light, and connect them to specific topological features. Further, our approach, characterizing local topological properties, could be used to investigate local effects of boundaries in finite size or experimental systems~\cite{Torquato2001}.  We remark that closing the gap with experimental systems would involve extending the analysis from point patterns to a broader class of HU patterns, such as HU heterogeneous media and scalar fields, which are also amenable to a topological analysis.

\section*{Acknowledgements}
We thank Marco Abbarchi for useful discussions and valuable comments on the manuscript. M.S. acknowledges support from the German Research Foundation (DFG) within the Research Training Group GRK 2868 D$^3$ - project number 493401063. D.J.S. acknowledges NSF Award DMS-1764421 with Simons Foundation grant 597491. A.V. acknowledges the German Research Foundation under Germany's Excellence Strategy, EXC-2068-390729961, Cluster of Excellence Physics of Life at TU Dresden. The authors also gratefully acknowledge computing time granted by the Center for Information Services and High-Performance Computing [Zentrum für Informationsdienste und Hochleistungsrechnen (ZIH)] at TU Dresden and the MIT SuperCloud and Lincoln Laboratory Supercomputing
Center.

\clearpage
\newpage

\setcounter{equation}{0}
\setcounter{figure}{0}
\setcounter{table}{0}
\setcounter{section}{0}
\setcounter{page}{1}

\renewcommand{\thesection}{S-\Roman{section}}
\renewcommand{\theequation}{S\arabic{equation}}
\renewcommand{\thefigure}{S\arabic{figure}}
\renewcommand{\bibnumfmt}[1]{[S#1]}
\renewcommand{\citenumfont}[1]{S#1}

\onecolumngrid

\begin{center}
  \textbf{\large \hspace{5pt} Supplemental Material \\ \vspace{0.05cm} Persistent homology and topological statistics of hyperuniform point clouds}\\[.2cm]
  
  \vspace{10pt}
  Marco Salvalaglio,$^{1,2}$, Dominic J. Skinner,$^{3}$
   J\"orn Dunkel$^{4}$, Axel Voigt$^{1,2,5}$ \\[.1cm]
  {\itshape \small
  ${}^1$Institute  of Scientific Computing,  Technische  Universit\"at  Dresden,  01062  Dresden,  Germany
  \\ 
  ${}^2$Dresden Center for Computational Materials Science (DCMS), TU Dresden, 01062 Dresden, Germany
 \\
  ${}^3$NSF-Simons Center for Quantitative Biology, Northwestern University,
2205 Tech Drive, Evanston, IL 60208, USA
 \\
  ${}^4$Department of Mathematics, Massachusetts Institute of Technology,\\
77 Massachusetts Avenue, Cambridge, MA 01239, USA 
\\
  ${}^5$Cluster of Excellence Physics of Life, Technische Universit\"at Dresden, 01062 Dresden, Germany
 }
\vspace{0.5cm}
\end{center}

This supplemental material contains:
\begin{itemize}
\item Additional persistent diagrams and $\tau_0$ (death times of connected components) distributions illustrated as in Fig.~4 in the main text -- \ref{sl1}. 
\item Parameters of the Skew-Normal distribution fitted on the distributions of $\tau_0$ -- \ref{sl2}. 
\item Distances $W_i(h_j^{\rm(I)},h_j^{\rm(II)})$ as in Fig.~5 in the main text for $i=1,2$ and $j=0,1$ -- \ref{sl3}. 
\item Residual variance as a function of embedding dimension quantifying the manifold dimensionality -- \ref{sl4}. 
\end{itemize}

\vspace{-0pt}
\section{Persistent Diagrams}
\label{sl1}

\begin{figure}[h!]
    \centering
    \includegraphics[width=\linewidth]{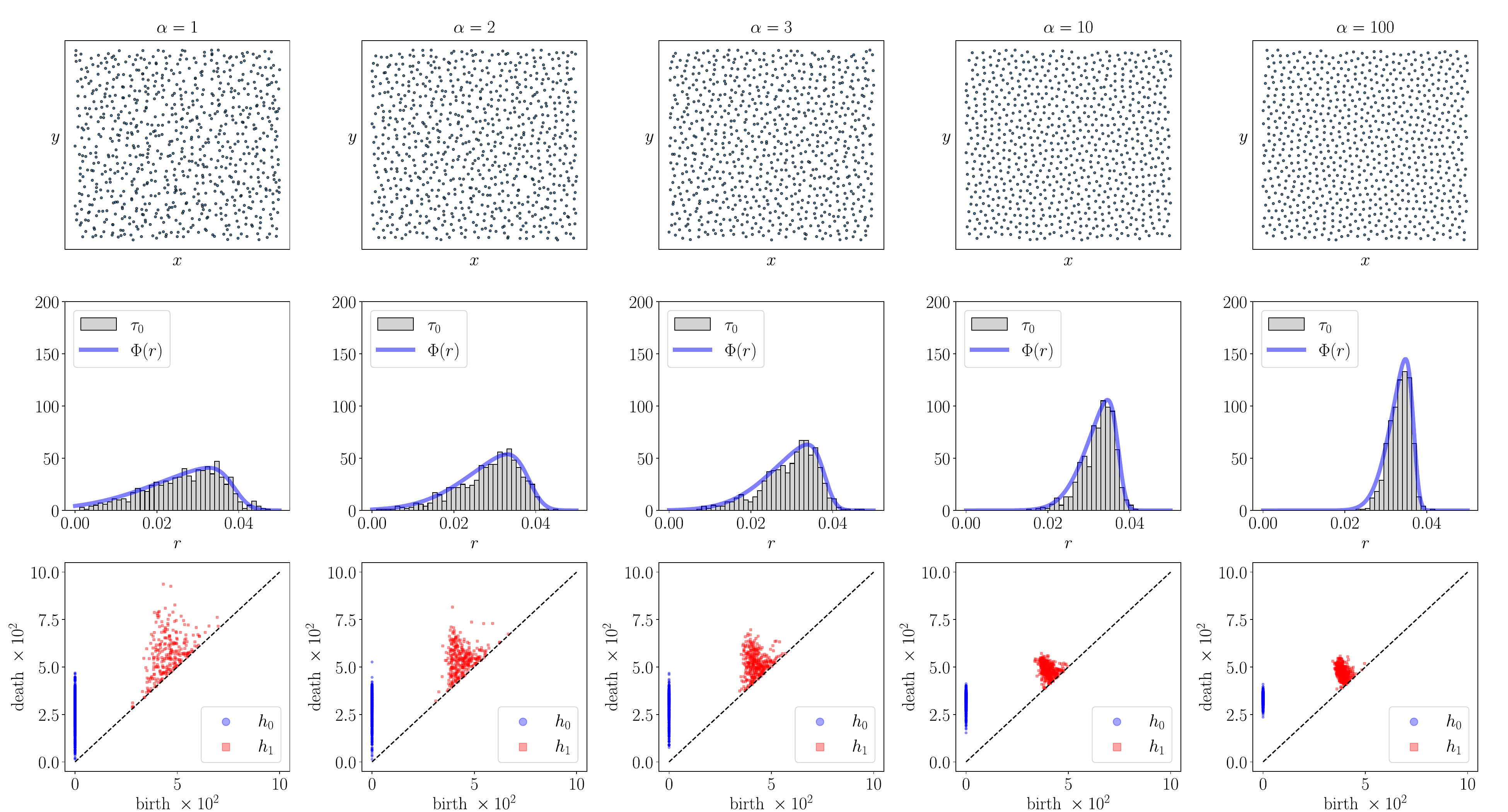}
    \caption{Point patterns (first row), distribution of $\tau_0$ (death times of connected components) as well as skew-normal distribution fit $\Phi(r)$ (second row), $h_0$ (connected components) and $h_1$ (holes) persistent diagrams (third row) varying $\alpha$ for $K=44\pi$ and $H=0.01$. See values of $\alpha$ in the first row.}
    \label{fig:figurePD1}
\end{figure}

\begin{figure}[h!]
    \centering
    \includegraphics[width=\linewidth]{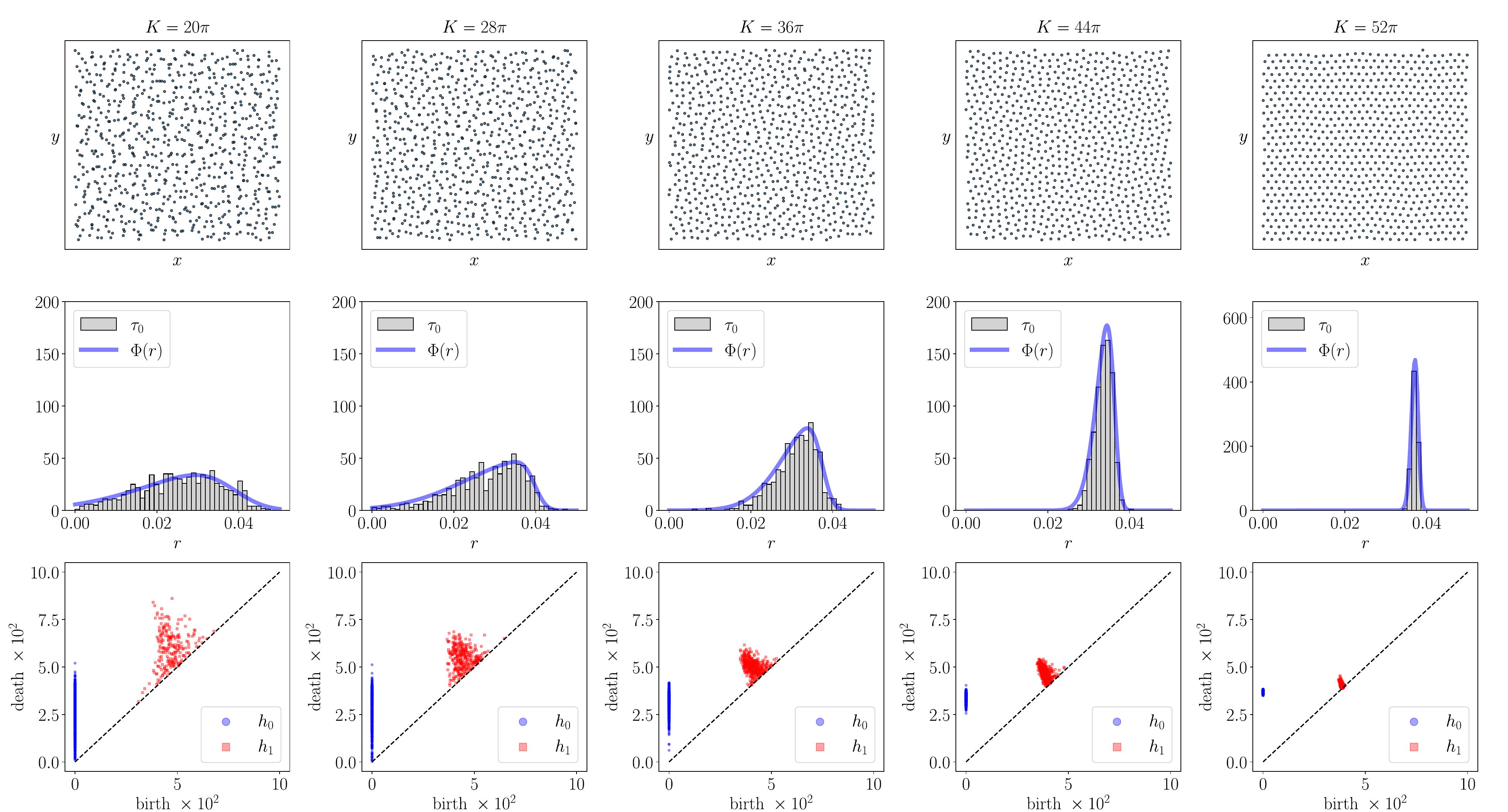}
        \caption{Point patterns (first row), distribution of $\tau_0$ (death times of connected components)as well as skew-normal distribution fit $\Phi(r)$ (second row), $h_0$ (connected components) and $h_1$ (holes) persistent diagrams (third row) varying $K$ for $\alpha=100$ and $H=10^{-4}$. See values of $K$ in the first row.}
    \label{fig:figurePD2}
\end{figure}

\begin{figure}[h!]
    \centering
    \includegraphics[width=\linewidth]{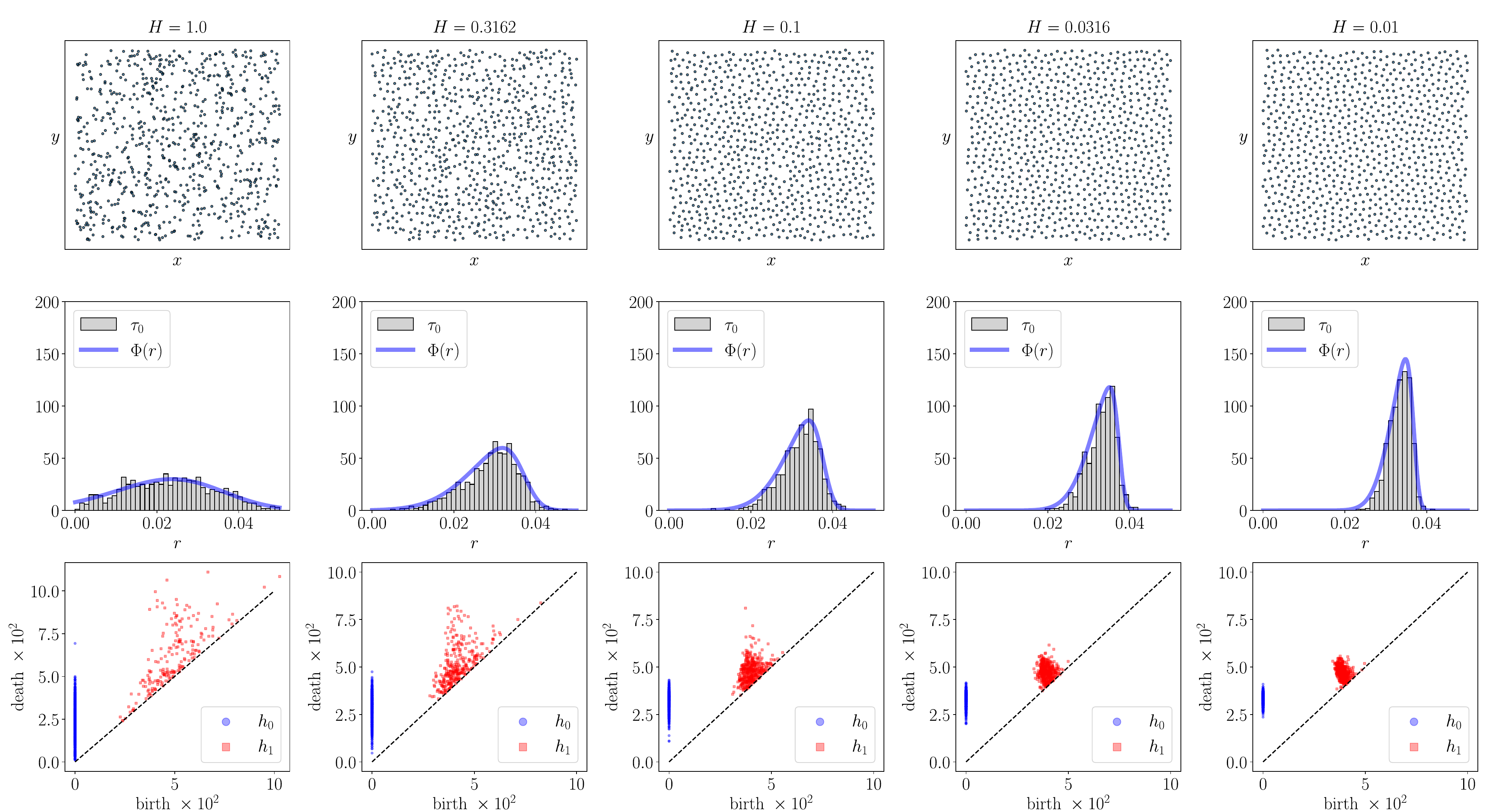}
    \caption{Point patterns (first row), distribution of $\tau_0$ (death times of connected components) as well as skew-normal distribution fit $\Phi(r)$ (second row), $h_0$ (connected components) and $h_1$ (holes) persistent diagrams (third row) varying $K$ for $\alpha=100$ and $K=44\pi$. See values of $H$ in the first row.}
    \label{fig:figurePD3}
\end{figure}

We report additional results complementing the analysis illustrated in Fig.~4 in the main text. We show point clouds obtained by varying $\alpha$, $K$, and $H$ (by keeping the other parameters constant) in Fig.~\ref{fig:figurePD1}, Fig.~\ref{fig:figurePD2}, and Fig.~\ref{fig:figurePD3}, respectively.


\newpage
\section{Parameters of the Skew-Normal distribution}
\label{sl2}

\begin{figure}[h!]
    \centering
    \includegraphics[width=\linewidth]{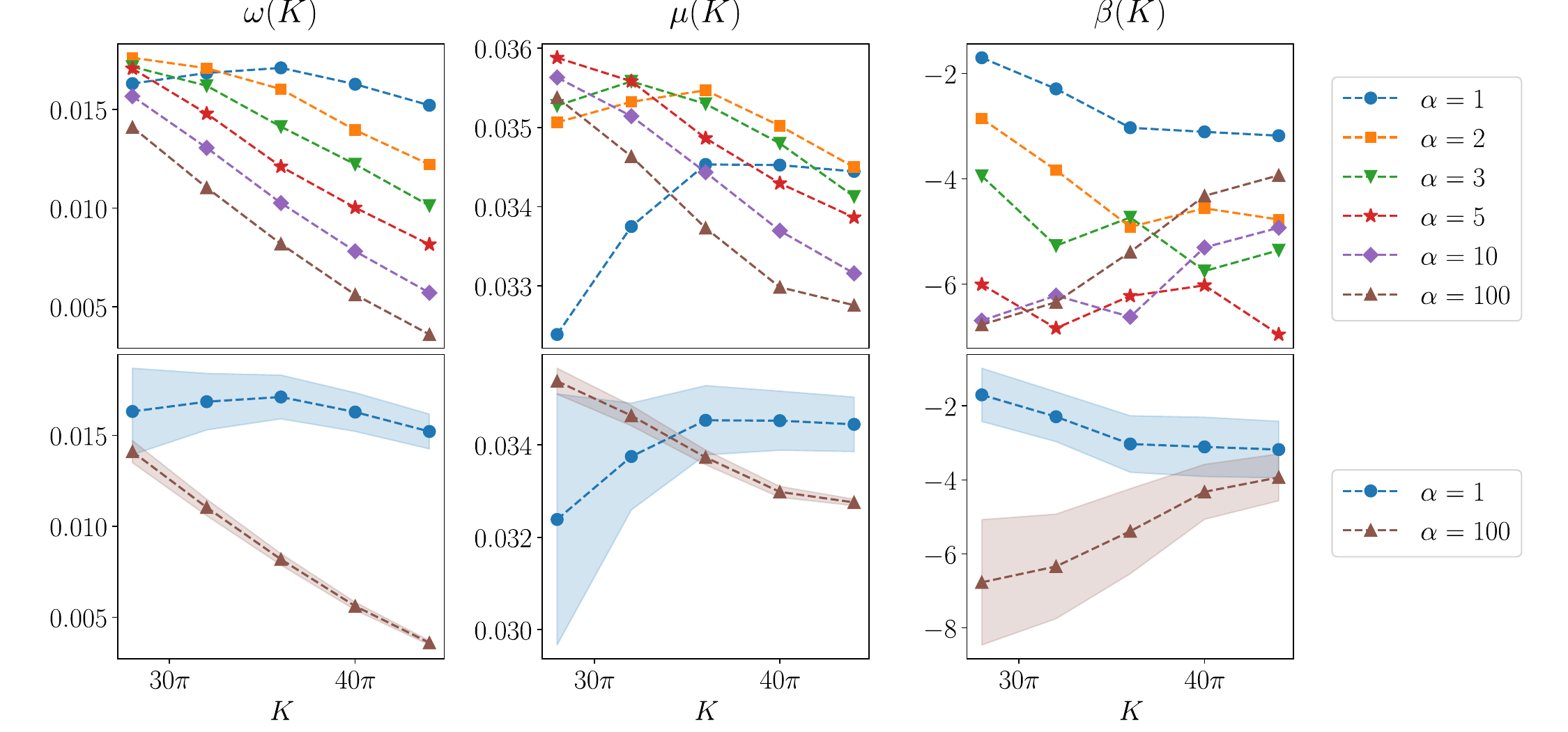}
    \caption{Parameters of $\Phi(r)$ fitted on the histograms $\tau_0$ by varying $K$ (on the x-axis) and $\alpha$ (different curves), $H=10^{-4}$. Shaded areas show the 68\% confidence intervals.}
    \label{fig:figureFITa}
\end{figure}

\begin{figure}[h!]
    \centering
    \includegraphics[width=\linewidth]{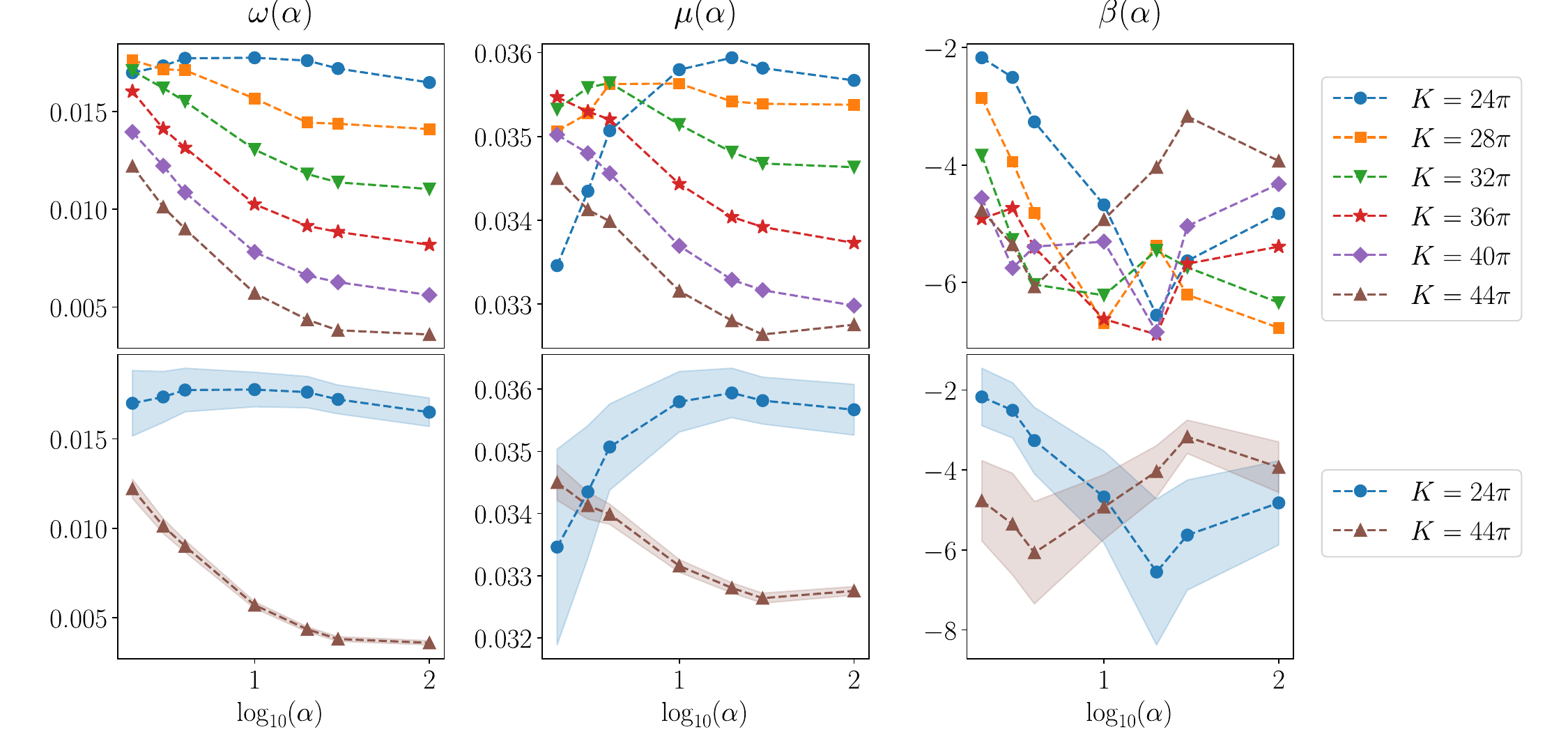}
    \caption{Parameters of $\Phi(r)$ fitted on the histograms $\tau_0$ by varying $\alpha$ (on the x-axis) for $K$ (different curves), $H=10^{-4}$. Shaded areas show the 68\% confidence intervals.}
    \label{fig:figureFITb}
\end{figure}

\begin{figure}[h!]
    \centering
    \includegraphics[width=\linewidth]{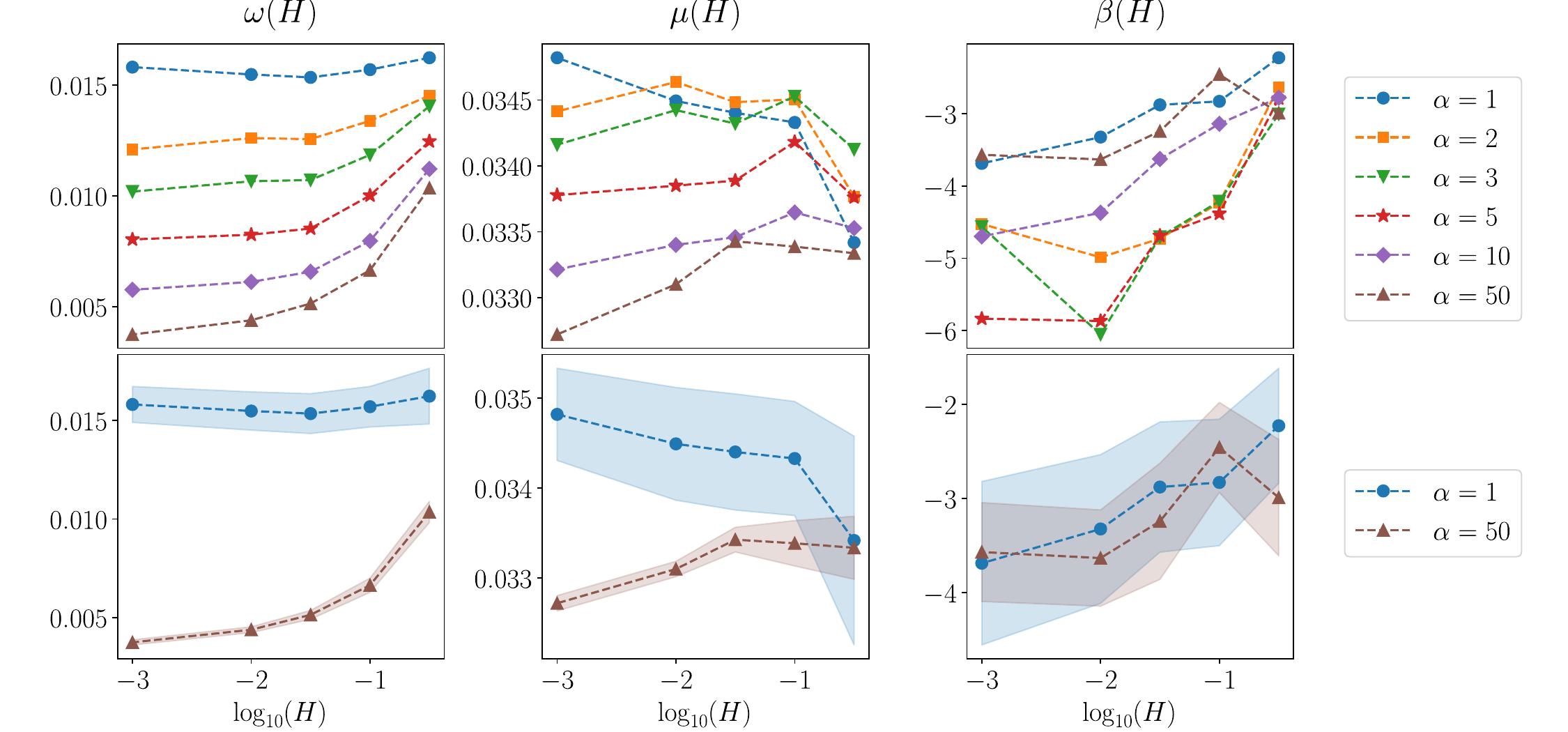}
    \caption{Parameters of $\Phi(r)$ fitted on the histograms $\tau_0$ by varying $H$ (on the x-axis) and $\alpha$ (different curves), $K=44\pi$. Shaded areas show the 68\% confidence intervals.}
    \label{fig:figureFITc}
\end{figure}

In the main text and Figs.~\ref{fig:figurePD1}--\ref{fig:figurePD3}, we have shown that the distribution of $\tau_0$ is well described by a skew-normal distribution
\begin{equation}\label{eq:skewnormal}
\Phi(r)=\frac{2}{\omega}\mathcal{N}\bigg(\frac{r-\mu}{\omega} \bigg)\mathcal{E}\bigg( \beta \frac{r-\mu}{\omega}\bigg),
\end{equation}
with 
\begin{equation}
\begin{split}
\mathcal{N}(x)&=\frac{1}{\sqrt{2\pi}}e^{-\frac{x^2}{2}},
\\
\mathcal{E}(x)&=\frac{1}{2}\bigg[1+\text{erf}\left(\frac{x}{\sqrt{2}} \right) \bigg].
\end{split}
\end{equation}
Fitting parameters $\omega$ (the width), $\mu$ (the location), and $\beta$ (the skewness) of the skew-normal distribution for representative patterns are fully illustrated here in Figs.~\ref{fig:figureFITa}--\ref{fig:figureFITc}. 
A negative skewness is observed for all the cases. We find $\omega$ decreases for increasing $\alpha$ and $K$ for relatively large values of these parameters, quantifying the increase in \textit{uniformity} corresponding to narrower distributions. However, for small $\alpha$, $\omega$ is non-monotonic in $K$ with a finite maximum. $\mu$ shows a similar trend, although the value of $\mu$ shows limited variation ($\sim 10\%$ across all analyzed cases) as we vary the parameters in the structure factor. The skewness $\beta$ exhibits an opposite behavior, namely a decrease (increase) of the negative skewness, hence an increase of the distribution asymmetry, for small (large) $\alpha$ and $K$. For marked HU characters, i.e., large $\alpha$ and $K$, the observed trend for the skewness results however rather noisy. This may be ascribed to the larger standard deviation accompanying the fitting of the skewness on narrower and narrower distributions.

\vspace{30pt}

\newpage

\vspace{30pt}

\vspace{30pt}

\newpage
\section{Distances between persistent diagrams}
\label{sl3}

\begin{figure}[h!]
    \centering
    \iflong
 \includegraphics[width=0.32\linewidth]{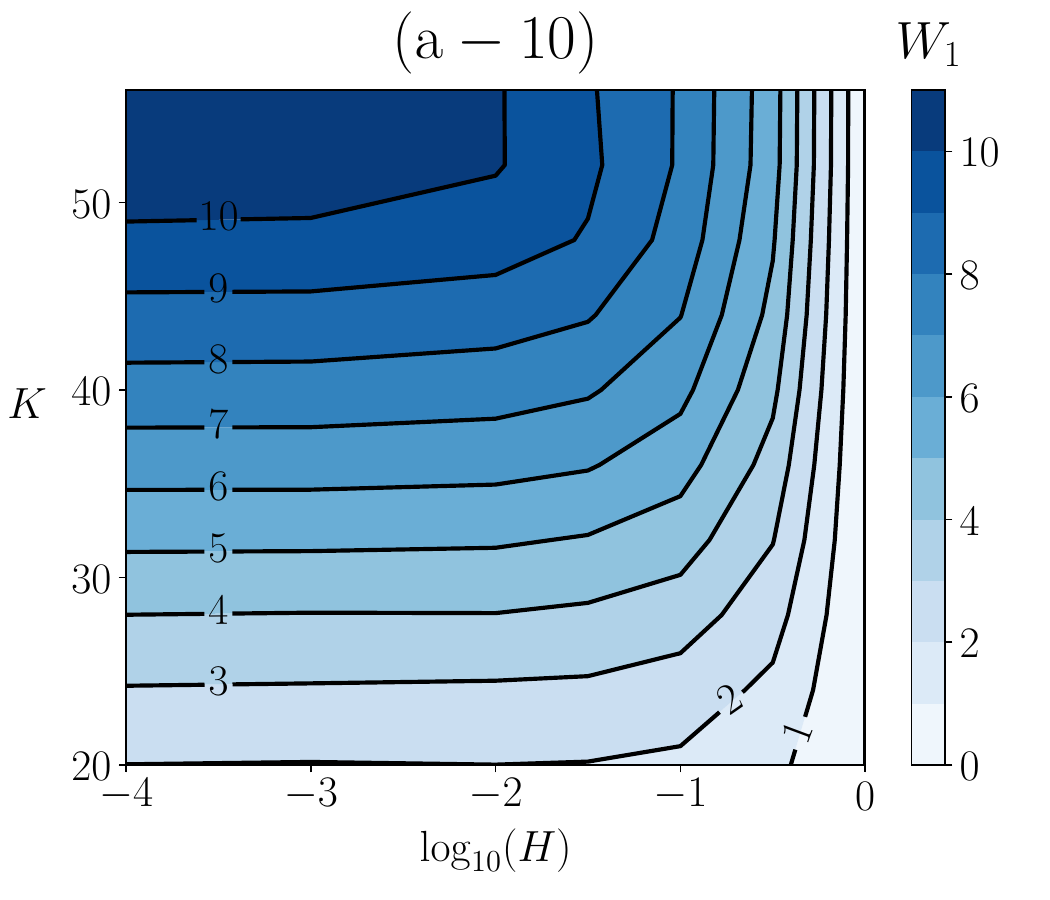}
    \includegraphics[width=0.32\linewidth]{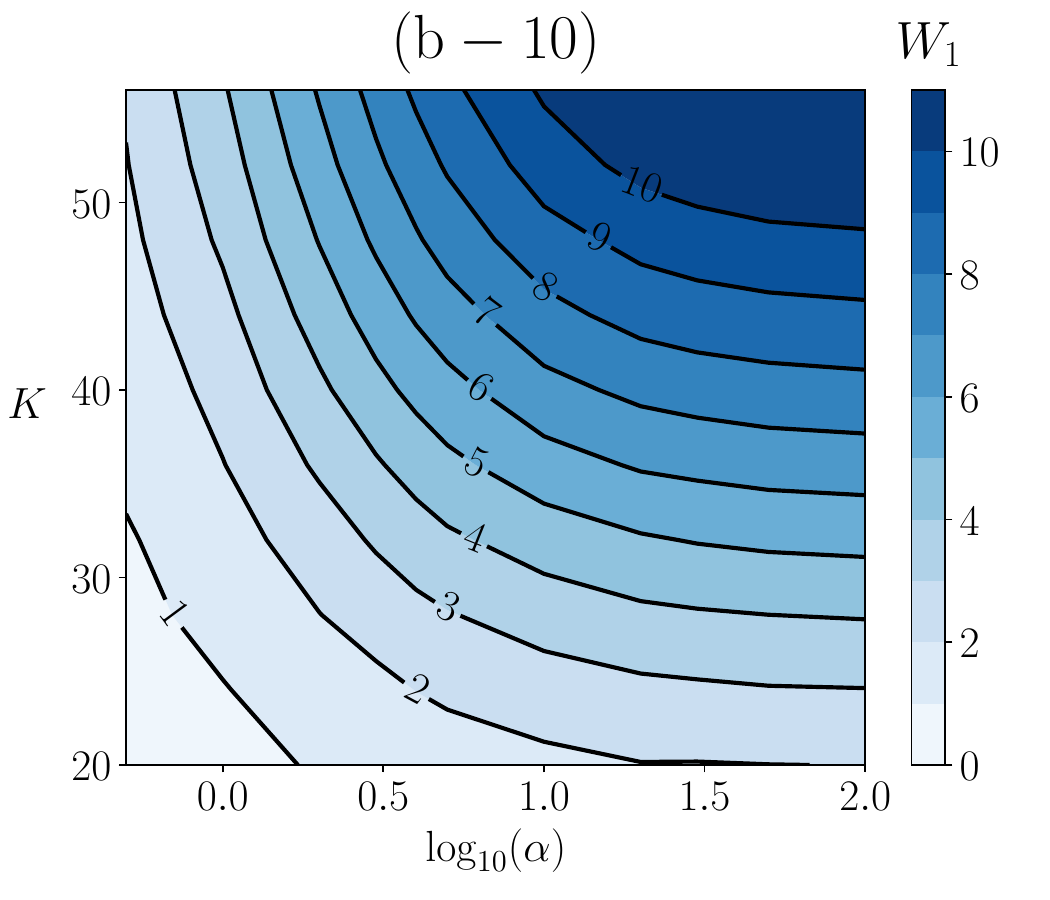}
    \includegraphics[width=0.32\linewidth]{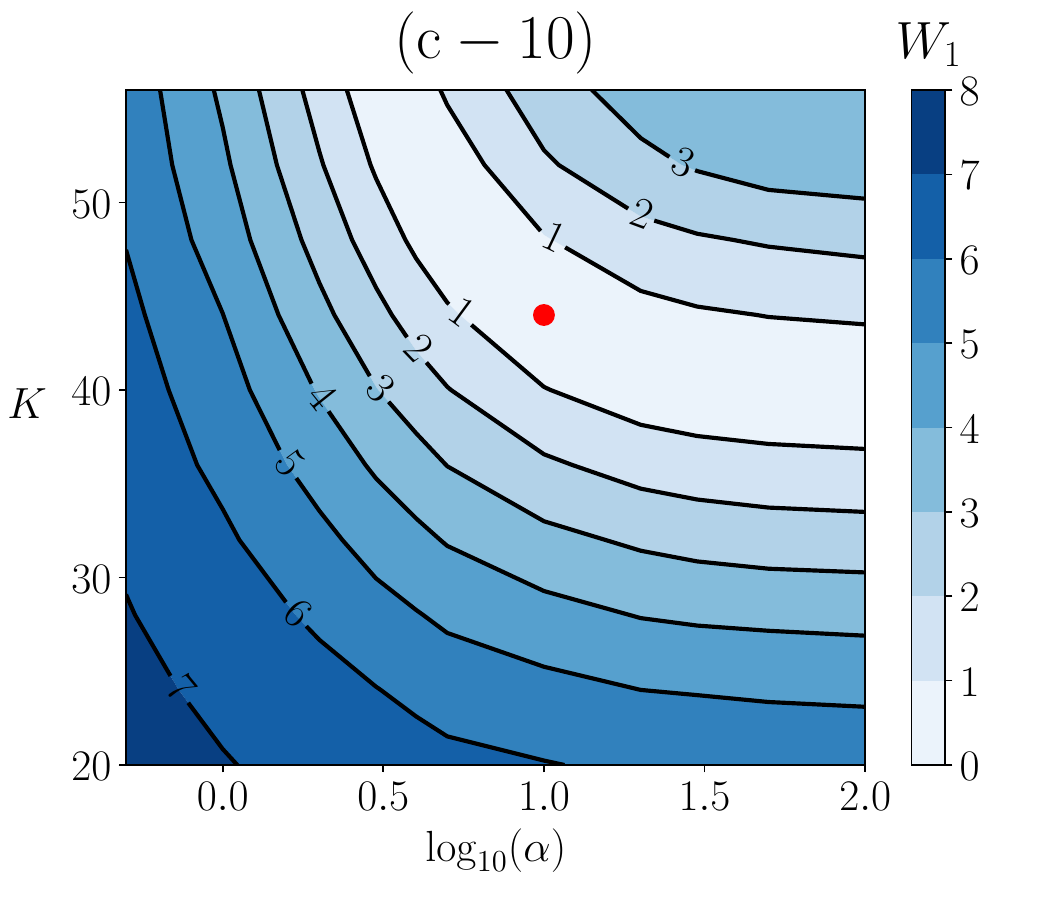}
    \\
    \includegraphics[width=0.32\linewidth]{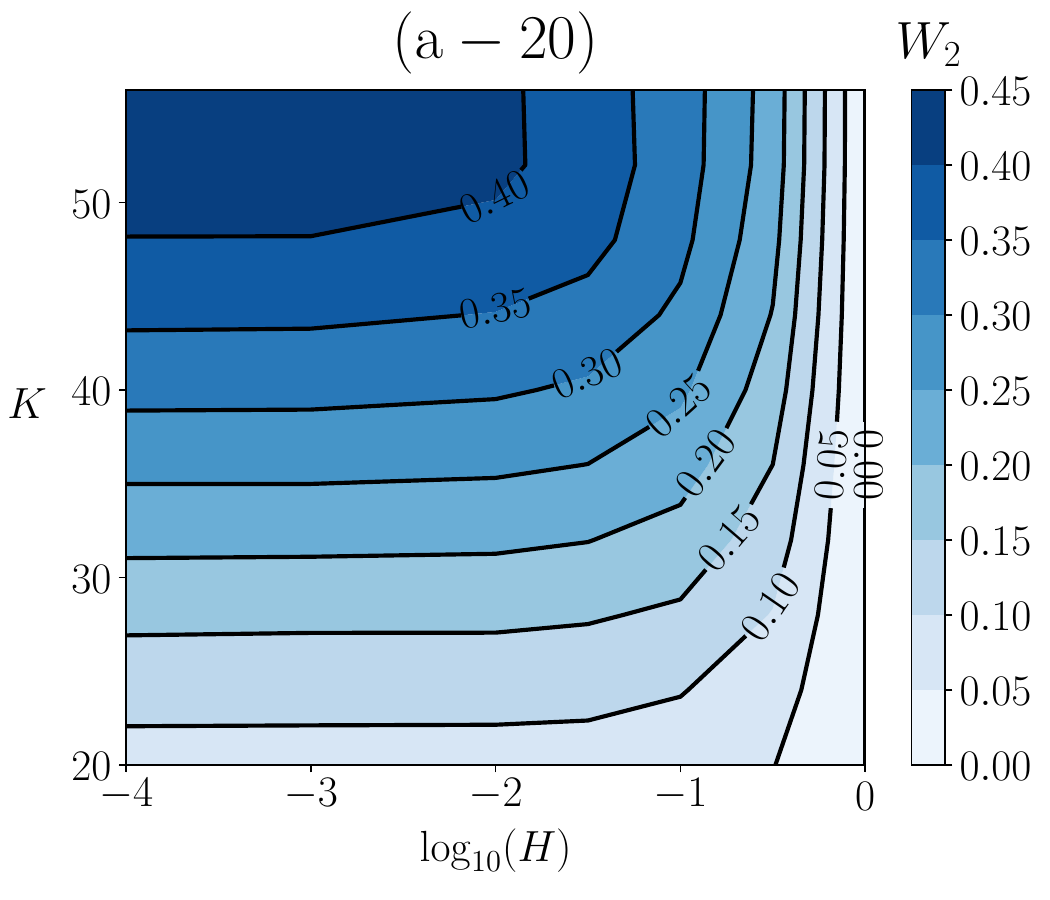}
    \includegraphics[width=0.32\linewidth]{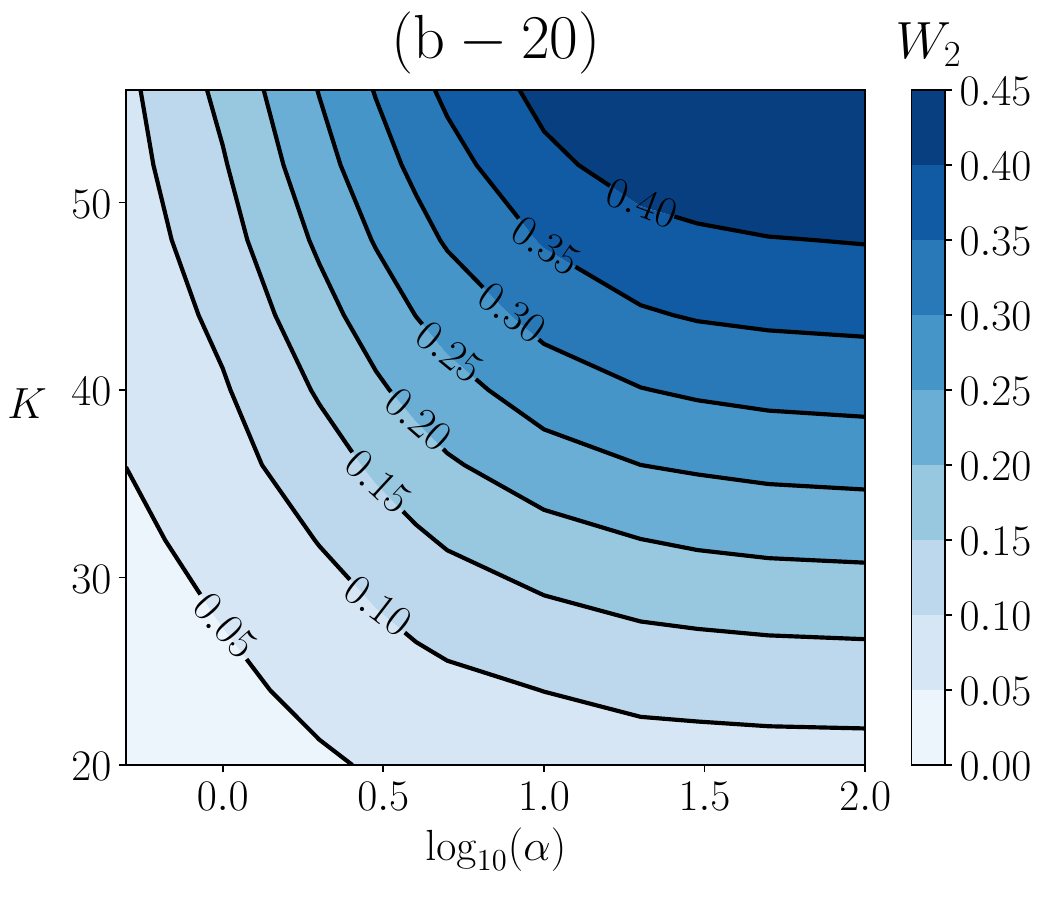}
    \includegraphics[width=0.32\linewidth]{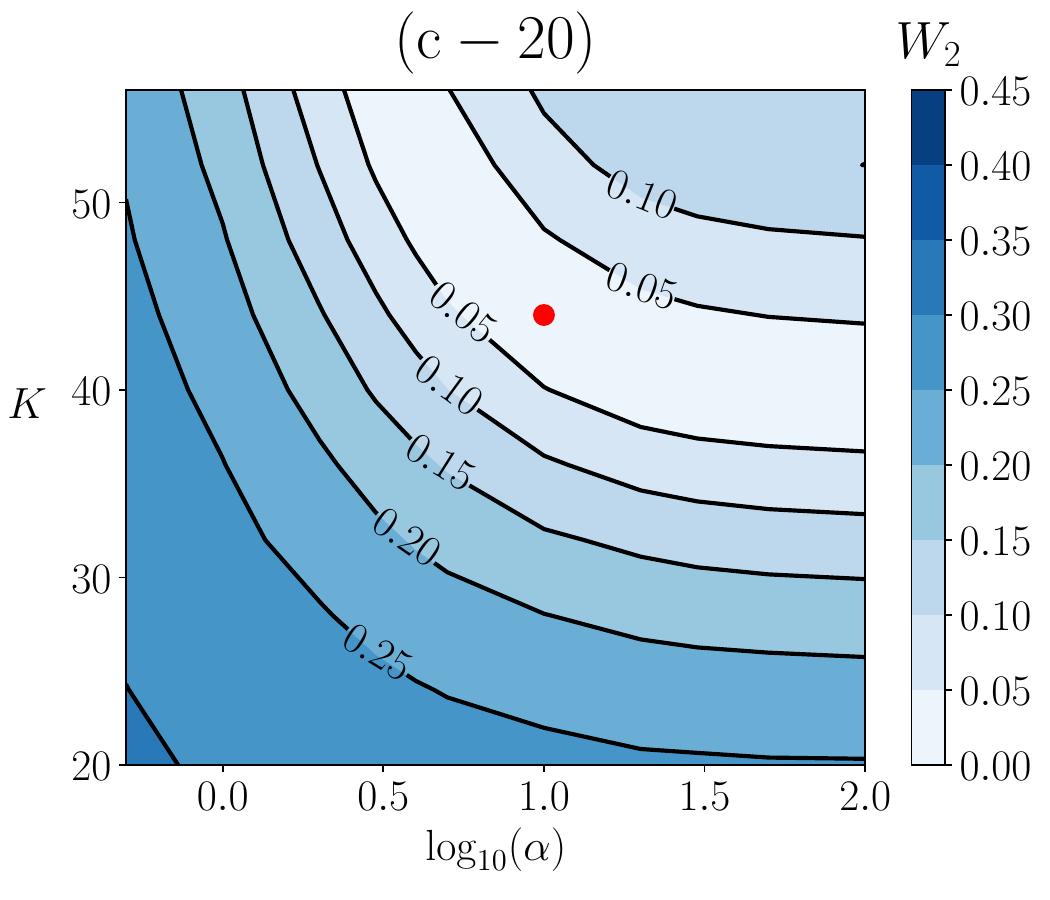}
\\
    \includegraphics[width=0.32\linewidth]{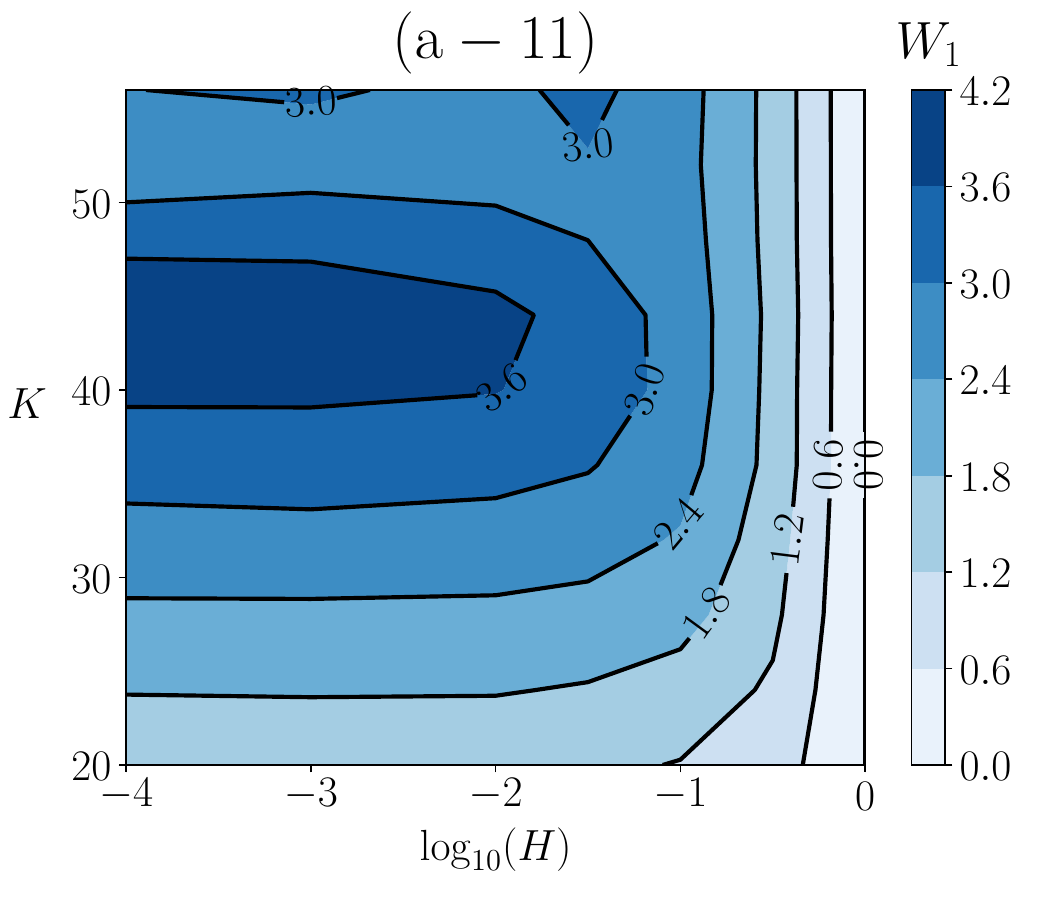}
    \includegraphics[width=0.32\linewidth]{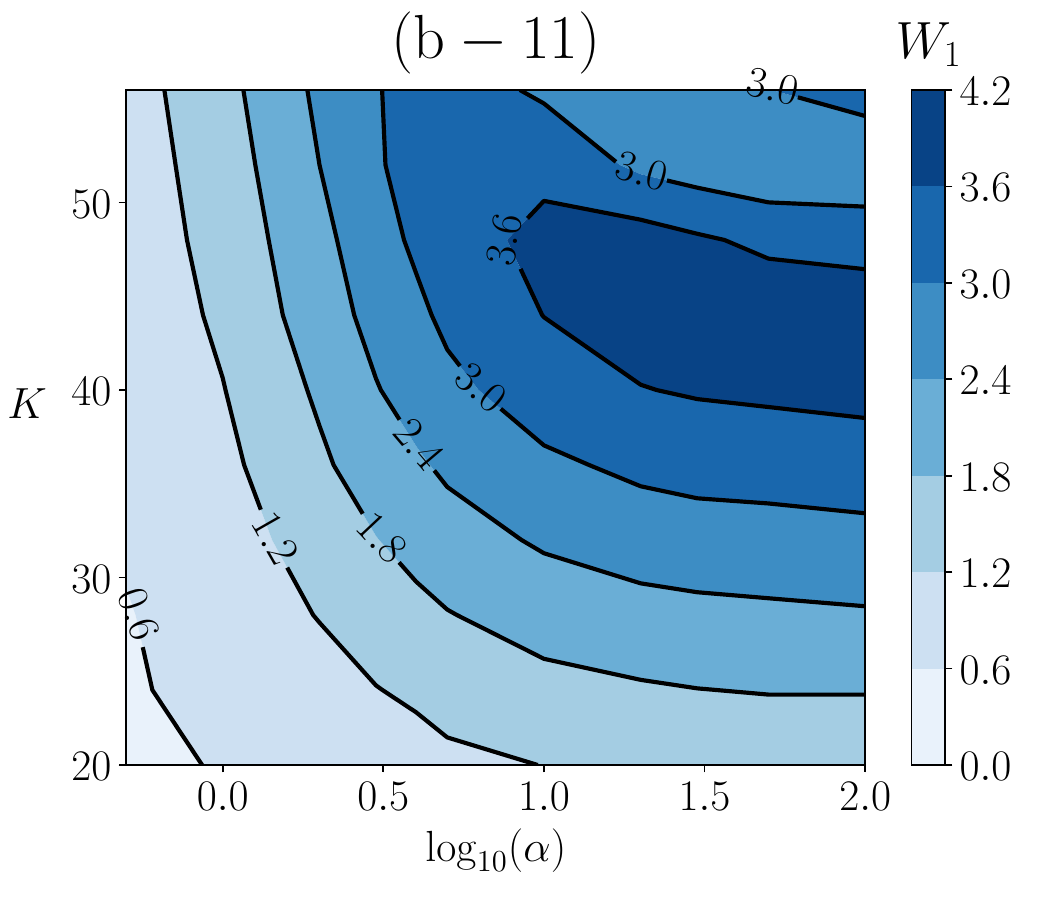}
    \includegraphics[width=0.32\linewidth]{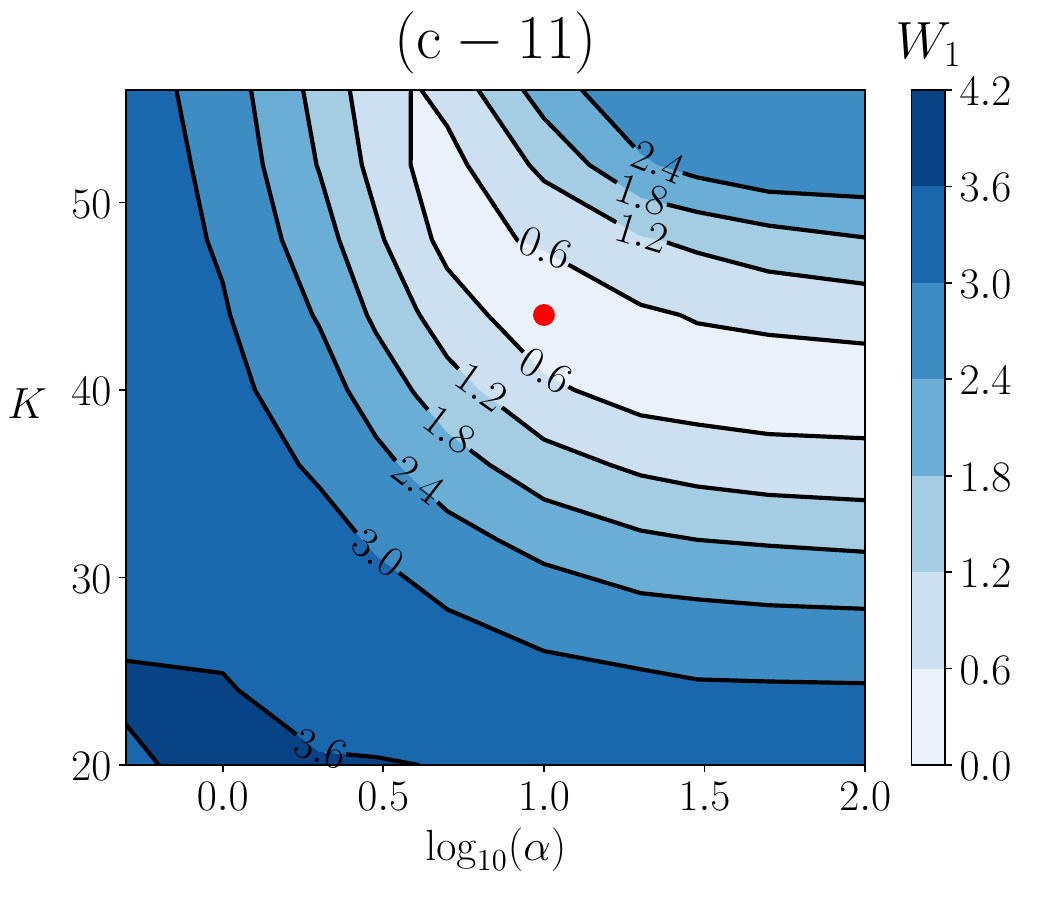}
\\
        \includegraphics[width=0.32\linewidth]{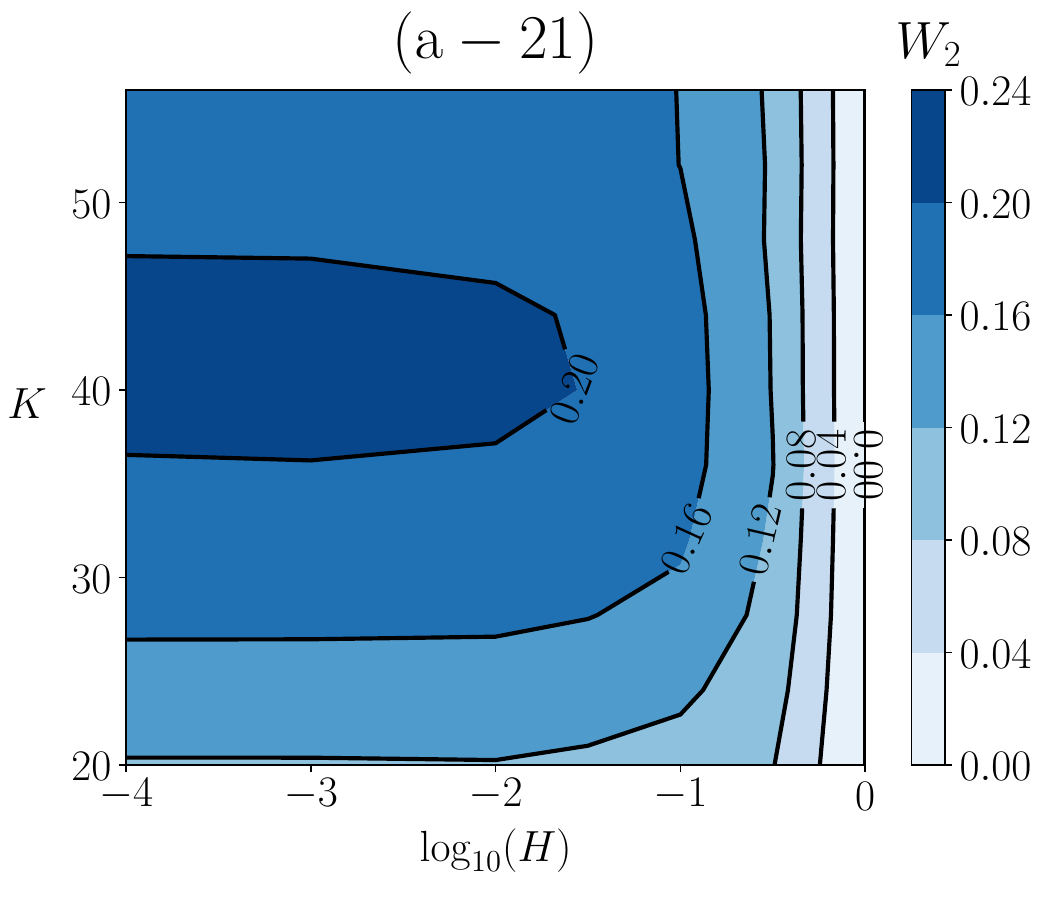}
    \includegraphics[width=0.32\linewidth]{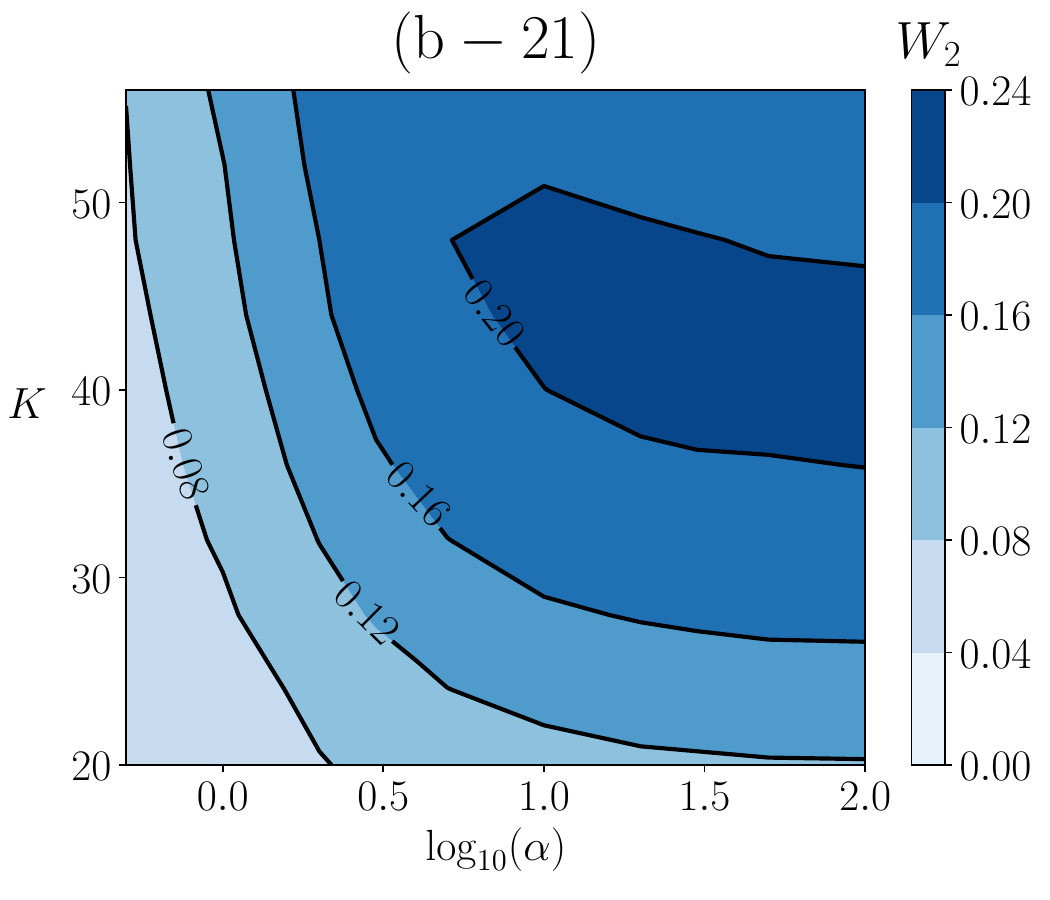}
    \includegraphics[width=0.32\linewidth]{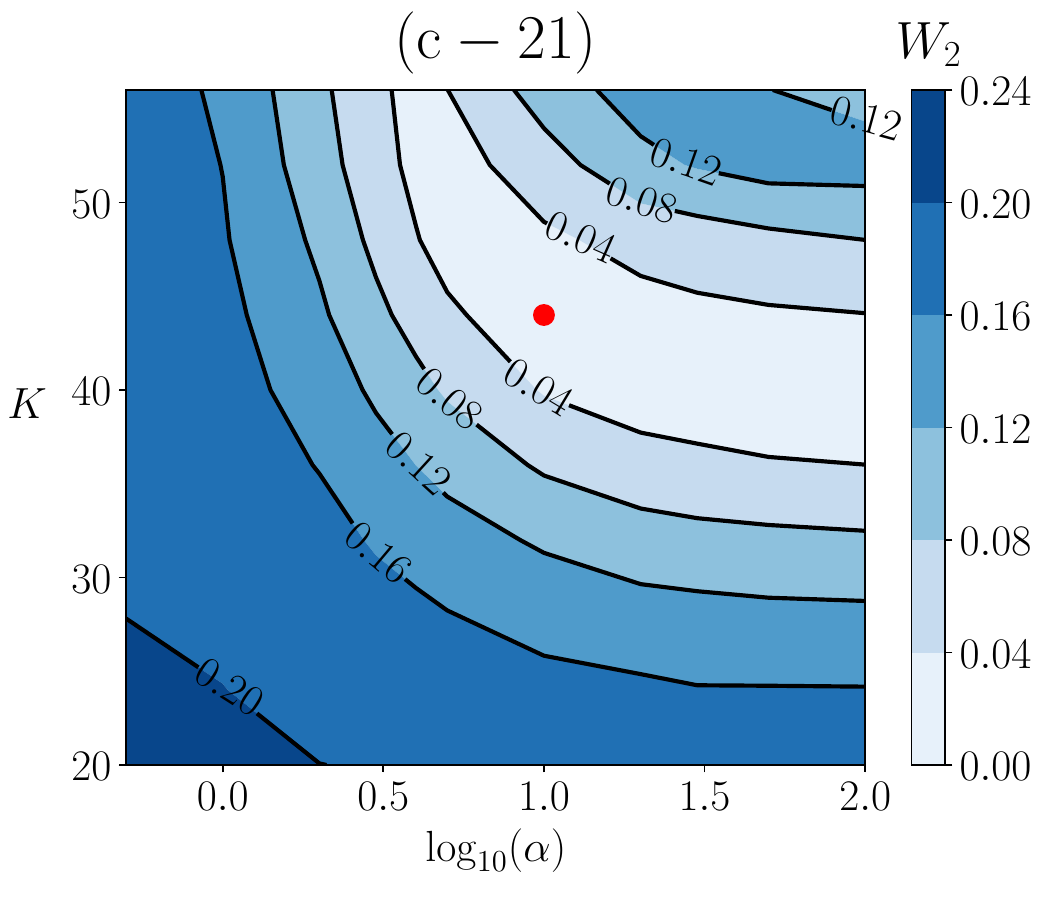}
    \fi
    \caption{Illustration of distances $W_i(h_j^{\rm (I)},h_j^{\rm (II)})$ as in Fig.~5 of the main text with $i=1,2$ and $j=0,1$. In all columns (I) is held fixed and (II) is varied. In the first and second column (a, b), ${\rm (I)}$ is a random arrangement. For ${\rm (II)}$ in the first column, (a), we vary $H$ and $K$ with $\alpha=100$ (Stealthy HU settings for sufficiently small $H$). For ${\rm (II)}$ in the second column, (b), we vary $\alpha$ and $K$ with $H=10^{-4}$. In the third column (c), ${\rm (I)}$ is a fixed HU arrangement with $K=44\pi$, $H=10^{-4}$, $\alpha=10$, whereas for ${\rm (II)}$,  we vary $\alpha$ and $K$ with $H=10^{-4}$. The parameter values corresponding to ${\rm (I)}$ are marked as a red dot. 
    }
    \label{fig:figureWDSI}
\end{figure}

\newpage
\section{Residual variance}
\label{sl4}

\begin{figure}[h!]
    \centering
    \includegraphics[width=\linewidth]{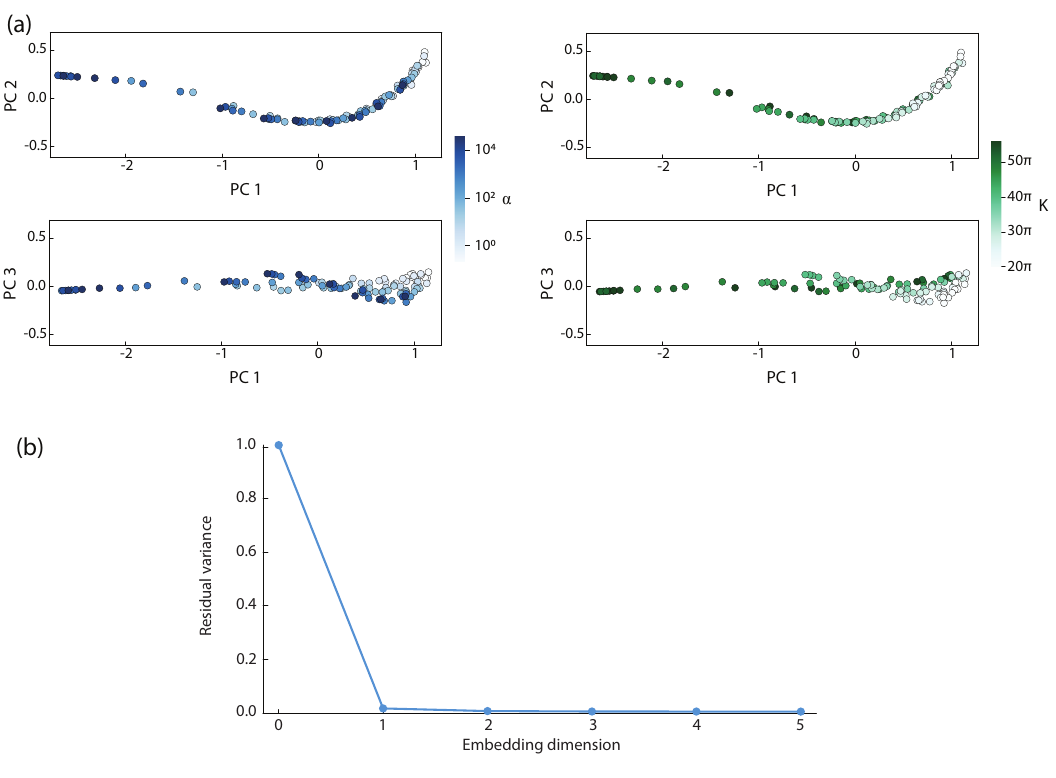}
    \caption{\label{fig:ResidualVariance}Quantifying the manifold dimension using the residual variance. (a) MDS embedding of all point patterns across various $K$ and $\alpha$ parameters, for the first three principal components (top is PC1 vs PC2, shown in main text Fig. 9, bottom is PC1 vs PC3). We can see that the manifold is primarily one dimensional and changing either $K$ or $\alpha$ primarily moves a point along the one dimensional curve. For more disordered systems (large PC 1 values), the manifold starts to become more two dimensional with $K$ and $\alpha$ being able to be independently varied. (b) To quantify the dimensionality, we plot the residual variance as a function of embedding dimension. Very little variance remains after a one dimensional embedding, showing the data manifold is essentially one dimensional. Residual variance is defined as $1 - R^2(\mathcal{D},\hat{D})$, where $\mathcal{D}$ is the true pairwise distance matrix, $\hat{D}$ is the Euclidean distance matrix in a low dimensional embedding, and $R^2$ is the standard correlation coefficient. Here, the low dimensional embedding is an isomap embedding rather than MDS, in order to ``unroll'' the manifold and find its intrinsic dimension, see Ref.~\cite{SISkinner2021} for details.}
\end{figure}

\end{document}